\documentclass[lettersize,journal]{IEEEtran}
\pdfoutput=1 
\usepackage{cite}
\usepackage{algorithm}
\usepackage{algpseudocode}
\usepackage{epsfig}
\usepackage{graphicx}
\usepackage{amsfonts}
\usepackage{comment}
\usepackage{caption}
\usepackage{color}
\usepackage{amssymb}
\usepackage{amsmath,dsfont,mathrsfs}
\usepackage{amsthm,mathtools}
\usepackage{multirow}
\usepackage{etoolbox}
\usepackage{balance}
\usepackage{booktabs}
\usepackage{enumitem}
\usepackage[table]{xcolor}

\usepackage[colorlinks = true,
    linkcolor = red,
    urlcolor  = black,
    citecolor = blue, 
    anchorcolor = magenta]{hyperref}

\usepackage{tikz}

\usepackage{caption}
\usepackage{boldline}
\usepackage{makecell}
\setcellgapes{1pt}
\usepackage{rotating}
\usepackage{lineno}

\AfterEndEnvironment{lemma}{\noindent\ignorespaces}

\newtheorem{definition}{Definition}[section]
\AfterEndEnvironment{definition}{\noindent\ignorespaces}

\AfterEndEnvironment{example}{\noindent\ignorespaces}

\ifCLASSINFOpdf
\else
\fi
\begin{document}
\title{TPMCF: Temporal QoS Prediction using Multi-Source Collaborative Features}

\author{Suraj~Kumar,
        Soumi~Chattopadhyay,~\IEEEmembership{Member,~IEEE,} and
        Chandranath~Adak,~\IEEEmembership{Senior Member,~IEEE} 

\IEEEcompsocitemizethanks{\IEEEcompsocthanksitem S. Kumar and S. Chattopadhyay are with the Dept. of CSE, Indian Institute of Technology, Indore, India-452020.  

C. Adak is with the Dept. of CSE, Indian Institute of Technology, Patna, India-801106.

\emph{Corresponding author:} S. Chattopadhyay (email: soumi@iiti.ac.in)

This work has been submitted to the IEEE for possible publication. Copyright may be transferred without notice, after which this version may no longer be accessible.
}

}


\maketitle

\begin{abstract}
The e-commerce industry has seen significant growth in recent years due to the introduction of new web service APIs. Quality-of-Service (QoS) parameters, which are fundamental for assessing service performance, have become crucial in evaluating services in the competitive market. Since QoS parameters can vary among users and change over time, accurate QoS predictions have become essential for users when selecting the most suitable services.
Existing methods for predicting temporal QoS have hardly achieved the desired accuracy, beset by challenges like data sparsity, the presence of anomalies, and the inability to capture intricate temporal user-service interactions. Although some recent approaches, particularly those founded on recurrent neural network-based sequential architectures, endeavor to model temporal relationships in QoS data, they grapple with performance degradation due to the omission of other pivotal features, such as collaborative relationships and spatial characteristics of users and services. Furthermore, the uniform attention among features across all time-steps can thwart progress in predictive accuracy.
This paper addresses these challenges and proffers a scalable strategy for temporal QoS prediction using multi-source collaborative features that not only furnishes heightened responsiveness but also engenders enhanced prediction accuracy. The method amalgamates collaborative features stemming from both users and services, capitalizing on the user-service relationship. Additionally, it integrates spatio-temporal auto-extracted features through the orchestration of graph convolution and a specialized variant of the transformer encoder equipped with multi-head self-attention. 
The proposed approach has been validated on the WSDREAM-2 benchmark datasets, and the results of these extensive experiments demonstrate that our framework surpasses major state-of-the-art methods in terms of predictive accuracy, all the while upholding robust scalability and reasonable responsiveness.
\end{abstract}

\begin{IEEEkeywords}
Temporal QoS Prediction, 
Graph Convolutional Matrix Factorization, 
Predictive Transformer Encoder
\end{IEEEkeywords}

\section{Introduction}\label{sec:intro}
\noindent
Over the years, businesses have increasingly turned to service-oriented architectures to improve user experiences and offer personalized recommendations \cite{soa}. With the rapid expansion of services, choosing an appropriate from a vast repository of functionally similar options has become resource-intensive and time-consuming, potentially incurring substantial costs. Consequently, delivering service recommendations that cater to individual user satisfaction has evolved into a fundamental research challenge in the domain of services computing \cite{Survey1-TSC_Zheng}.

A common practice is to recommend services based on Quality-of-Service (QoS) parameters, as these parameters (e.g., response time, throughput) are commonly employed to evaluate web service performance. However, it is important to recognize that QoS parameters are not static for a service; they exhibit variations among users and can even fluctuate over time for a single user. Consequently, accurate QoS prediction becomes a crucial prerequisite for making service recommendations, which is the primary focus of this paper.

Numerous contemporary methods primarily concentrate on addressing QoS prediction across users but often overlook the critical temporal dependencies existing among QoS values \cite{OFFDQ,DAFR-2020}. The long-established approaches used for the static QoS prediction are based on collaborative filtering (CF) \cite{Survey2-TSC}, which are classified into three categories: 
(a) memory-based approaches using similarity between users and services \cite{UPCC,IPCC,WSRec}, 
(b) model-based approaches using matrix factorization (MF) \cite{NMF,NIMF}/ factorization machine \cite{EFM}/ deep-architectures \cite{DAFR-2020}, and 
(c) hybrid methods combining both memory and model-based approaches \cite{CAHPHF,OFFDQ}. 
In general, these approaches explicitly leverage various features, including user and service similarities, statistical parameters derived from QoS logs, and contextual information like geographical and network-related attributes. These features span from low-level to high-level characteristics. However, it is worth noting that these methods are designed to predict QoS values while often overlooking the temporal context of QoS data. Static QoS prediction methods are apparently insufficient for making accurate QoS predictions because most of the QoS parameters fluctuate over time, even for a specific user-service pair. 

Recent developments in QoS prediction methods have started to address the temporal dimension. Temporal smoothing (TS) \cite{TUIPCC,TF-KMP,TMF} has been a popular technique, but it often falls short in effectively utilizing temporal features, leading to decreased prediction accuracy. To overcome this, an alternative approach employing the ARIMA model \cite{TASR} has been proposed, although it comes with computational costs and reduced interpretability compared to TS.
Tensor factorization (TF) \cite{WSPRED,CTF,BNLFT,NNCP,WLRTF} models have been introduced to leverage the triadic relationship among users, services, and time. However, these models may struggle to capture the complex, high-dimensional features in QoS data sequences.
In recent studies, deep learning models, including Long Short-Term Memory (LSTM) \cite{PLMF,RTF,OPST,Mul-TSFL,QSPC} and Gated Recurrent Unit (GRU) \cite{DeepTSQP,RNCF}, have shown potential for improving prediction accuracy over traditional methods. Nevertheless, the inability to capture intricate, higher-order triadic relationships among user-service interactions over time due to limitations in feature representation can hinder the performance of these deep learning-based methods.
Most deep learning-based approaches in the literature rely on either QoS features derived from the temporal QoS invocation log matrix \cite{TF-KMP,TUIPCC} or contextual features like geographical location and network parameters (e.g., IP address, autonomous system) for predicting QoS values \cite{Mul-TSFL,QSPC}. However, there is a recognized need for a more comprehensive exploration of user-service interactions to enhance prediction accuracy.

In light of the aforementioned limitations within the existing literature, this paper introduces TPMCF, a novel framework. TPMCF harnesses the power of Graph Convolution \cite{GCN} in conjunction with a Transformer Encoder \cite{TRANSFORMER} to effectively utilize spatio-temporal collaborative features, enabling it to capture the triadic relationships embedded within temporal QoS invocation sequences. TPMCF offers a dual advantage: (a) it excels at capturing multi-source collaborative features, enhancing the accuracy of temporal QoS prediction, (b) it leverages a transformer encoder with multi-head attention to effectively capture temporal dependencies among QoS data. We now summarize our \textbf{contributions}. 

\emph{(i) Utilization of multi-source collaborative features for QoS prediction}: 
TPMCF effectively utilizes collaborative features from both users and services to capture intricate, higher-order relationships among user-service interactions over time, achieving enhanced prediction accuracy. 
On the one hand, TPMCF combines explicit features derived using domain knowledge with auto-extracted features for the target user-service pair. On the other hand, TPMCF auto-extracts spatial features by exploring neighborhood relationships and temporal features by analyzing the user-service interactions over time.
  
\emph{(ii) Spatial feature extraction using graph convolution}: 
We propose a graph convolutional matrix factorization (GCMF) module to auto extract spatial features by investigating the neighborhood of the given user-service pair. On the one hand, GCMF addresses the issue of data sparsity, a prevalent issue in the realm of temporal QoS prediction \cite{Survey1-TSC_Zheng}. On the other hand, GCMF excels at capturing complex relationships within QoS data, thereby enhancing the accuracy of QoS prediction.

\emph{(iii) QoS prediction using a transformer encoder}: 
We further propose a temporal QoS prediction module comprising a transformer encoder followed by a fully connected neural network. In general, the transformer encoder works better than LSTM \cite{OPST} and GRU \cite{DeepTSQP}-based sequential architectures due to its self-attention mechanism \cite{TRANSFORMER}, as observed empirically. In this paper, we adopt the transformer encoder architecture \cite{TRANSFORMER} and propose an improvised version, namely, the predictive transformer encoder (PTE),  responsible for capturing the temporal dependencies among QoS data.  
The PTE takes a sequence of the spatial features for a given user-service pair, extracted by the GCMFs and generates spatio-temporal features by exploiting the temporal dependencies among the feature sequences, which is the earliest attempt of its kind. 
A fully connected neural network finally uses the spatio-temporal features for prediction.
    
\emph{(iv) Extensive experiments}: 
We extensively experimented on two benchmark datasets of WSDREAM-2 \cite{WSDREAM}. Experimental results indicate that our framework outperformed major state-of-the-art approaches in terms of prediction accuracy.

The rest of the paper is organized as follows. Section \ref{sec:problem} presents an overview of the problem formulation. Section \ref{sec:method} then discusses the proposed framework in detail. The experimental results are analyzed in Section \ref{sec:results}, while the literature review is presented in Section \ref{sec:related_work}. Finally, Section \ref{sec:conclusion} concludes this paper.


\section{Problem Formulation}\label{sec:problem}
\begin{figure}[!h]
    \centering
    \includegraphics[width=0.6\linewidth]{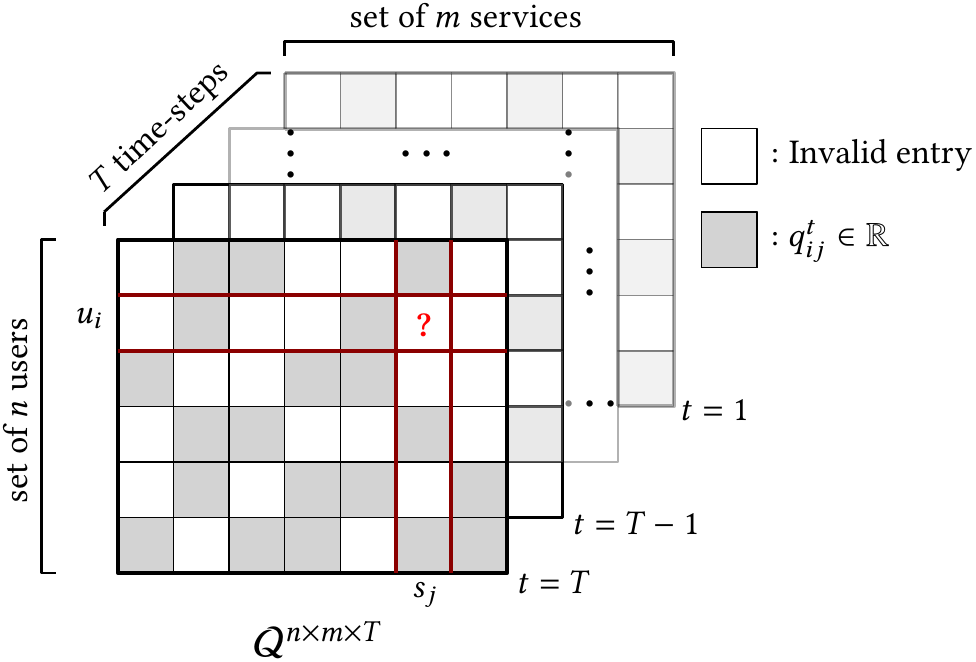}
    \caption{Temporal QoS prediction}
    \label{fig:problem}
\end{figure}

\noindent
The pictorial overview of the temporal QoS prediction problem is presented in Fig. \ref{fig:problem}, which is mathematically formulated in this section. 
Given the following inputs to the framework:
\begin{itemize}
    \item A set of $n$ users ${\mathcal{U}} = \{u_1, u_2, \ldots, u_n\}$
    \item A set of $m$ services ${\mathcal{S}} = \{s_1, s_2, \ldots, s_m\}$
    \item A QoS parameter $q$
    \item A set of observations on $q$ for user-service pair $(u_i, s_j) \in {\mathcal{U}} \times {\mathcal{S}}$ for {past $T$ time-steps}
    \item A QoS invocation log ${\mathcal{Q}}$ in the form of a tensor with dimension $n \times m \times T$, as defined below:
    $$
        {\mathcal{Q}}(i, j, t) = 
        \begin{cases}
            \multirow{2}{*}{$q_{ij}^t \in \mathbb{R}^+$} & \text{value of } q \text{ of } s_j \text{ invoked by }  u_i \\  & \text{at $t^{th}$ time-step} \\
            \multirow{2}{*}{0} & \text{Otherwise } \\ & \text{(representing invalid entry)}
        \end{cases}
    $$
    \noindent
\end{itemize}
The objective is to predict the value of $q$ for a target user-service pair $(u_i, s_j)$ at the time-step $T$.

\begin{figure*}[!t]
    \centering
    \includegraphics[width=0.8\linewidth]{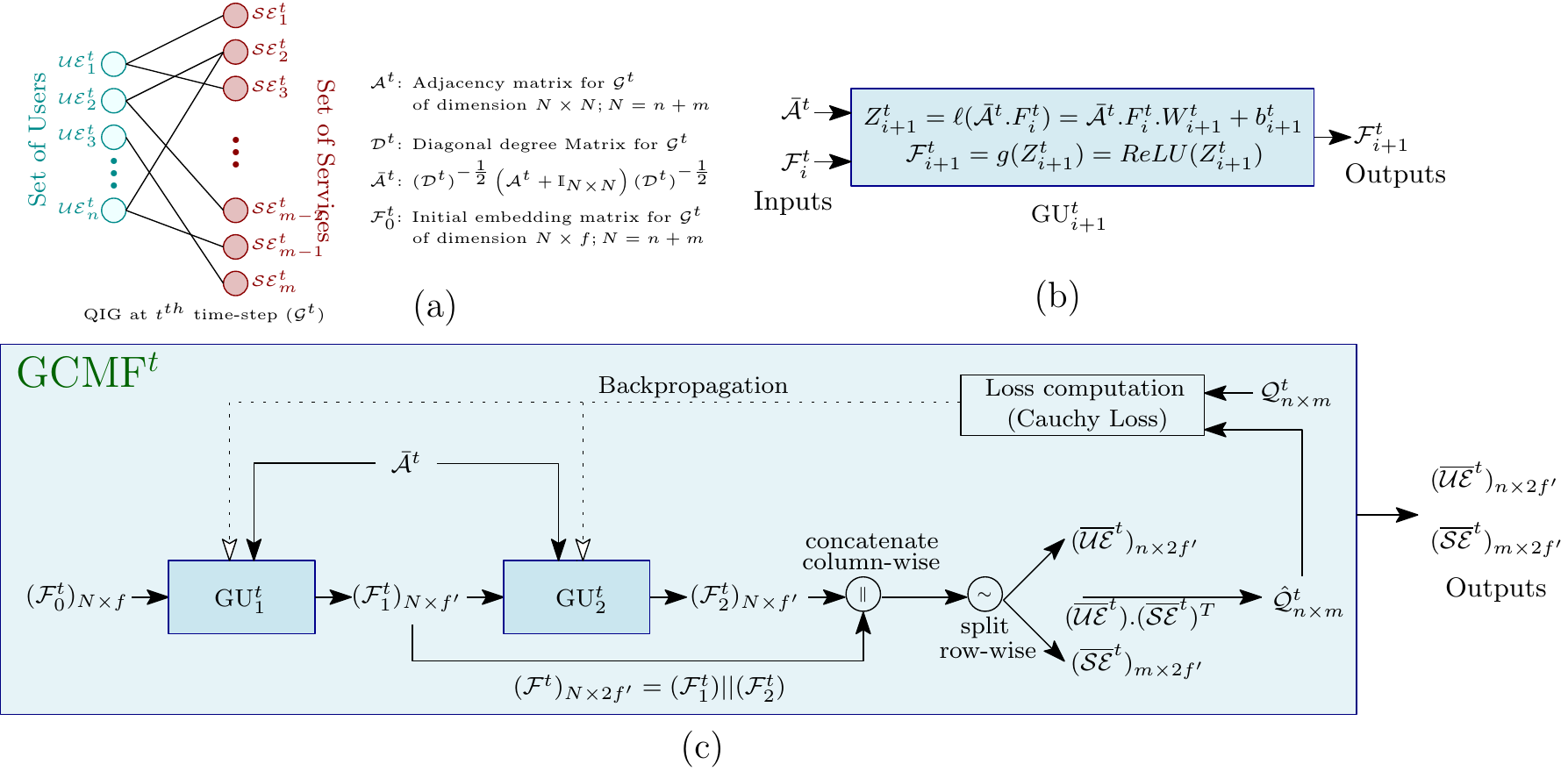}
    \caption{Collaborative spatial feature extraction module (CSFE) using graph convolutional matrix factorization (GCMF)}
    \label{fig:gcn}
\end{figure*}

\section{TPMCF: Proposed Framework}
\label{sec:method}
\noindent
This section discusses our proposed framework TPMCF for temporal QoS prediction, which has two primary modules: 
(a) \emph{Collaborative spatial feature extraction module (CSFE)}: responsible for extracting spatial features using graph convolutional matrix factorization (GCMF), and 
(b) \emph{Temporal QoS prediction module (TQP)}: accountable for generating spatio-temporal features utilizing a variant of transformer encoder, followed by QoS prediction using a fully connected neural network. 
Given initial feature vectors of a user-service pair $u_i, s_j$ at a specific time-step, GCMF exploits the neighborhood of $u_i$ and $s_j$, and extracts spatial features that PTE then employs to generate spatio-temporal features to be used by a fully connected neural network for QoS prediction. Other than spatial feature extraction to enable the framework to achieve a high prediction accuracy, GCMF has another significant advantage when employed prior to the transformer encoder. GCMF handles the data-sparsity issues quite well. Therefore, we do not require any additional data imputation method to predict the missing values, which could impede achieving high prediction accuracy when any deep neural architecture does the estimation. We now explain each module of TPMCF in detail. We begin by describing the QoS invocation graph representation, which is one of the primary components the CSFE module of TPMCF.

\subsection{Representation of QoS Invocation Graph}
\noindent
We first define the QoS invocation graph (QIG).
\begin{definition}[QoS Invocation Graph (QIG)]\label{def:qig}
A QoS invocation graph ${\mathcal{G}}^t = (U^t \cup S^t, E^t, {\mathcal{UE}}^t \cup {\mathcal{SE}}^t)$ is a bipartite graph, where the vertices $U^t$ and $S^t$ represent the set of users ${\mathcal{U}}$ and the set of services $\mathcal{S}$, respectively. An edge $e_{ij}^t = (u^t_i, s^t_j) \in E^t$ exists between two vertices, $u^t_i \in U^t$ and $s^t_j \in S^t$, if ${\mathcal{Q}}(i, j, t) = q_{ij}^t \neq 0$. Each node $u^t_i \in U^t$ and $s^t_j \in S^t$ are associated with node vectors ${\mathcal{UE}}^t_i \in {\mathcal{UE}}^t$ and ${\mathcal{SE}}^t_j \in {\mathcal{SE}}^t$ representing the initial feature embedding of $u_i$ and $s_j$ at time-step $t$, respectively.
\hfill$\blacksquare$
\end{definition}
\noindent
An example of a QIG at time-step $t$ is shown in Fig. \ref{fig:gcn}(a). In CSFE module, a QIG ${\mathcal{G}}^t$ is represented by an adjacency matrix ${\mathcal{A}}^t$ of dimension $N \times N$, where $N = n + m$, as defined below.
\begin{equation}\small
 {\mathcal{A}}^t = (a_{ij}^t) \in \{0, 1\}^{N \times N};  a_{ij}^t = 
    \begin{cases}
        1, & \text{if } e^t_{ij} \in E^t \text{ of QIG } {\mathcal{G}}^t\\
        0, & \text{otherwise}
    \end{cases}
\end{equation}

\noindent
The adjacency matrix ${\mathcal{A}}^t$ helps to explore the neighborhood of each node to generate the spatial feature embedding for all the users and services. However, instead of using ${\mathcal{A}}^t$ directly in the CSFE module, we normalize ${\mathcal{A}}^t$ (denoted by ${\bar{\mathcal{A}}}^t$) as follows. 
\begin{equation}\label{eq:normalized_adjacency}\small
    {\bar{\mathcal{A}}}^t = ({\mathcal{D}}^t)^{-\frac{1}{2}}({\mathcal{A}}^t + \mathds{I}_{N \times N})({\mathcal{D}}^t)^{-\frac{1}{2}}
\end{equation}
where, ${\mathcal{D}}^t$ is a diagonal matrix representing the degree of each node of ${\mathcal{G}}^t$ by its diagonal elements, as defined below.
\begin{equation}\small
 {\mathcal{D}}^t = (d_{ij}^t) \in \mathbb{Z}_+^{N \times N}; \quad d_{ij}^t = 
    \begin{cases}
    1 + \sum \limits_{k = 1}^{N} {\mathcal{A}}^t(i, k), & \text{if } i = j\\
    0, & \text{otherwise}
    \end{cases}
\end{equation}

\noindent
Normalization is required here to avoid the more significant influence of the higher-degree nodes during learning \cite{GCN}. This is because the higher degree nodes do not necessarily mean they contribute more to the prediction. It only means they have more interactions at that time-step. Moreover, normalization helps in scaling.
In Eq. \ref{eq:normalized_adjacency}, an identity matrix (i.e., $\mathds{I}_{N \times N}$) is added to take into account the self-influence of a node while extracting the spatial features for it. 
It may be noted, each non-zero element of ${\bar{\mathcal{A}}}^t$, i.e., ${\bar{\mathcal{A}}}^t(i, j)$, is normalized by the square root of the number of invocations of the corresponding user $u_i$ and service $s_j$ at time-step $t$ as recorded in ${\mathcal{Q}}$, refers to Eq. \ref{eq:normalized_adjacency_element}.
\begin{equation}\label{eq:normalized_adjacency_element}\small
    {\bar{\mathcal{A}}}^t(i, j) = {{\mathcal{A}}^t(i, j)}/ \left({\sqrt{d^t_{ii}}.\sqrt{d^t_{jj}}}\right)
\end{equation}
${\bar{\mathcal{A}}}^t$ is finally used for the convolution operation in the CSFE module, which is discussed later in this section.

As mentioned in Definition \ref{def:qig}, each node in $(U^t \cup S^t)$ is associated with a node vector representing the feature embedding of the corresponding user or service. We now discuss the construction of the initial feature embedding for the users/services below.

\subsubsection{Construction of Initial Feature Embedding}\label{subsubsec:features}
\noindent
The initial feature embedding of a user or service comprises three distinct types of feature vectors, derived from ${\mathcal{Q}}$, as elaborated below.

    {\emph{(i) Statistical Features (${\mathcal{F}}_S^t$)}}: For each user $u_i \in {\mathcal{U}}$ or service $s_j \in {\mathcal{S}}$, we obtain the statistical features ${\mathcal{F}}_S^t(u_i)$ or ${\mathcal{F}}_S^t(s_j)$ consisting of 5 statistical parameters: minimum, maximum, median, mean, and standard deviation from their QoS invocation profile ${\mathcal{Q}}^t(u_i)$ or ${\mathcal{Q}}^t(s_j)$, where ${\mathcal{Q}}^t(u_i) = {\mathcal{Q}}(i, .~, t)$ and ${\mathcal{Q}}^t(s_j) = {\mathcal{Q}}(.~, j, t)$ represent the QoS invocation vectors of $u_i$ and $s_j$ at time-step $t$, respectively. Statistical features of a user or service capture the global characteristics of the invocation pattern of that user or service.
    
    {\emph{(ii) QoS Features (${\mathcal{F}}_Q^t$)}}: We use matrix decomposition \cite{NMF} of ${\mathcal{Q}}^t$ (i.e., QoS invocation log matrix at $t$) to obtain the QoS feature embedding of length $f_q$ for each user and service (say, ${\mathcal{F}}_Q^t(u_i)$ and ${\mathcal{F}}_Q^t(s_j)$) at $t$.
    
    
    {\emph{(iii) Collaborative features using correlations (${\mathcal{F}}_C^t$)}}: For each user $u_i \in {\mathcal{U}}$ or service $s_i \in {\mathcal{S}}$, we first obtain the correlation between $u_i$ or $s_i$ and every other user $u_j \in {\mathcal{U}}$ or service $s_j \in {\mathcal{S}}$ in terms of their QoS invocation profile ${\mathcal{Q}}^t(u_i)$ or ${\mathcal{Q}}^t(s_i)$ using cosine similarity \cite{COSINE-SIMILARITY}. It may be noted that for each $u_i$ or $s_i$, we have a vector of length $n$ or $m$, which is then fed to a stacked autoencoder \cite{STACKED-AUTOENCODER} to obtain the latent representation of correlation feature embedding of length $f_c$ for each user and service (say, ${\mathcal{F}}_C^t(u_i)$ or ${\mathcal{F}}_C^t(s_j)$) at $t$.

The final feature embedding for a user $u_i$ or service $s_j$ is constructed by concatenating the above three feature vectors.

\begin{minipage}{0.23\textwidth}\scriptsize
\begin{equation}\scriptsize
 {\mathcal{UE}}^t_i = {\mathcal{F}}_S^t(u_i) || {\mathcal{F}}_Q^t(u_i) || {\mathcal{F}}_C^t(u_i);
\end{equation}
\end{minipage}
\begin{minipage}{0.23\textwidth}\scriptsize
\begin{equation}\scriptsize
 {\mathcal{SE}}^t_j = {\mathcal{F}}_S^t(s_j) || {\mathcal{F}}_Q^t(s_j) || {\mathcal{F}}_C^t(s_j)
\end{equation}
\end{minipage}

\noindent
The initial feature embedding matrix ${\mathcal{F}}_0^t$ is of dimension $N \times f$, where $f = 5 + f_q + f_c$. The first $n$ rows of ${\mathcal{F}}_0^t$ contain user-feature embedding for $n$ users, while the last $m$ rows of ${\mathcal{F}}_0^t$ comprise service-feature embedding for $m$ services. 
This embedding matrix ${\mathcal{F}}_0^t$ is used in the CSFE module to incorporate neighborhood information for every user/service.

\subsection{Collaborative Spatial Feature Extraction (CSFE) Module using GCMF}
\noindent
We now discuss the automated feature generation using GCMF at time-step $t$. 
Fig. \ref{fig:gcn}(b) shows the details of a GCMF-unit (GU), which is the core component of the CSFE module. Given the normalized adjacency matrix ${\bar{\mathcal{A}}}^t$ and feature embedding matrix ${\mathcal{F}}^t_i$ to the $i^{th}$ GCMF-unit (GU$^t_i$), for each node $v_j^t \in U^t \cup S^t$ of ${\mathcal{G}}^t$, GU$^t_i$ combines the feature of $v_j^t$ with the features of all the nodes reachable from $v_j$ with a path length less than or equal to $i$ to obtain spatial feature vector for $v_j^t$.

In our CSFE module, two GUs are connected sequentially. 
To avoid the over-smoothing problem \cite{GRL-BOOK}, the output of GU$^t_1$ (i.e., ${\mathcal{F}}_1^t$) is concatenated with the output of GU$^t_2$ (i.e., ${\mathcal{F}}_2^t$) to obtain a matrix ${\mathcal{F}}^t$ of size $N \times 2f'$. 
Finally, ${\mathcal{F}}^t$ is split row-wise to generate user and service embedding matrices, 
${\overline{\mathcal{UE}}}^t$ and ${\overline{\mathcal{SE}}}^t$, respectively. 
Fig. \ref{fig:gcn}(c) presents the overview of the GCMF architecture. 
The output of the CSFE module is forwarded to the subsequent module of our framework.  

\begin{figure*}
    \centering
    \includegraphics[width=0.8\linewidth]{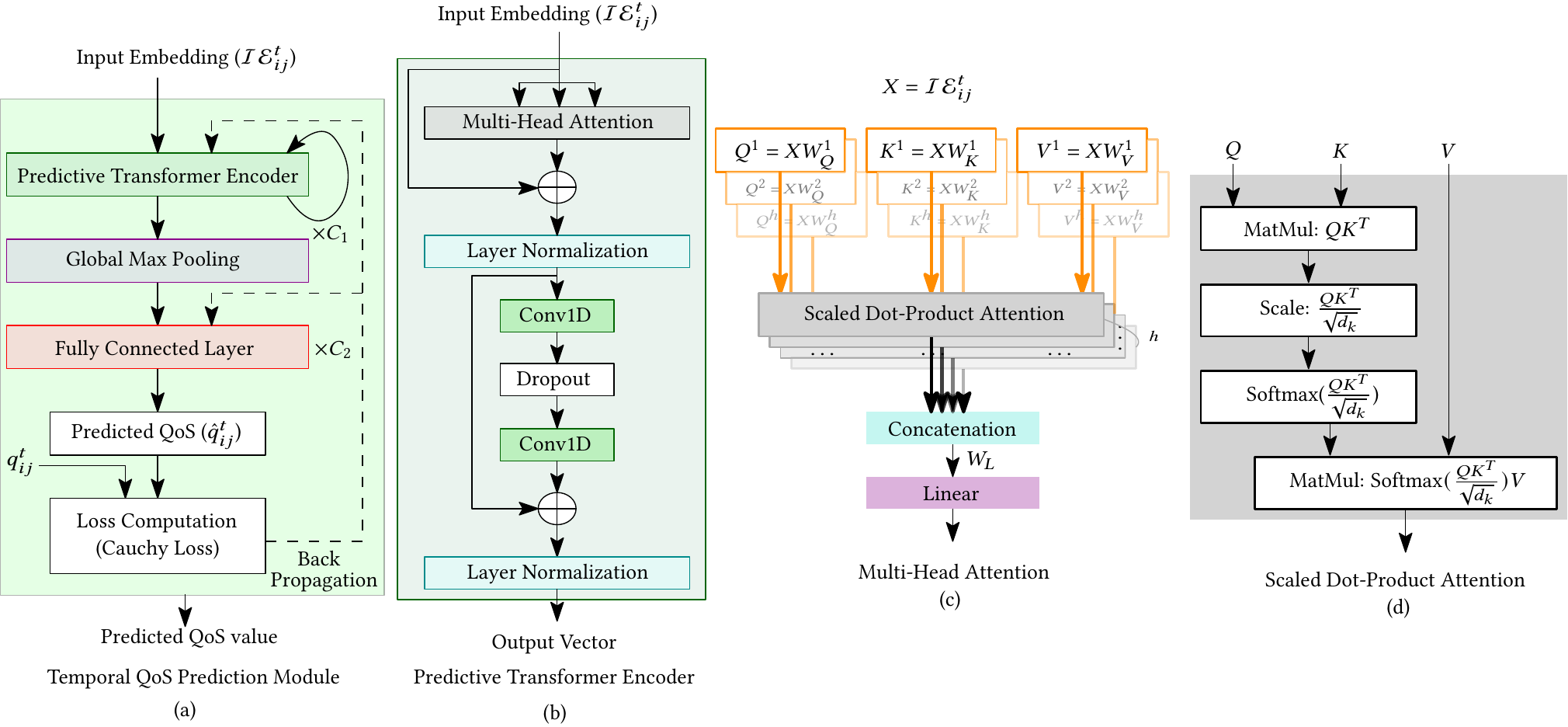}
    \caption{Temporal QoS Prediction Module (TQP) using Predictive Transformer Encoder (PTE)}
    \label{fig:transformer}
\end{figure*}

\subsection{Temporal QoS Prediction (TQP) Module using Predictive Transformer Encoder (PTE)}\label{subsec:te}
\noindent
This is the final module of our framework.
Given a target user-service pair $u_i$ and $s_j$, the objective of this module is to predict the QoS value for $u_i, s_j$ at time-step $t$. The main crux of this module is to predict the value of ${\mathcal{Q}}(i,j,k)$ while considering the temporal dependencies for the last $\mathcal{T}$ time-steps. 

The first step of this module is to generate the input embedding, which is then fed to a series of transformer encoder blocks, followed by global max-pooling and fully connected (FC) layers to predict the value of ${\mathcal{Q}}(i,j,k)$.
The pictorial overview of this module is shown in Fig. \ref{fig:transformer}(a). 
We now discuss each step in detail.

\subsubsection{Input Embedding Construction} 
At first, for each time-step $k$ (where, $ t-\mathcal{T}+1 \le k \le t$), the user embedding ${\overline{\mathcal{UE}}}_i^k \in {\overline{\mathcal{UE}}}^k$ and the service embedding ${\overline{\mathcal{SE}}}_j^k \in {\overline{\mathcal{SE}}}^k$ obtained from the CSFE module are concatenated to produce a vector ${\overline{\mathcal{E}}}_{ij}^k$ of length $4f'$. 
Finally, ${\overline{\mathcal{E}}}_{ij}^k$ across all $\mathcal{T}$ time-steps are concatenated row-wise to generate the input embedding ${\mathcal{IE}}_{ij}^t$ of dimension ${\mathcal{T}} \times 4f'$. 
\begin{equation}\scriptsize
 {\mathcal{IE}}_{ij}^t = (\underbrace{({\overline{\mathcal{UE}}}^{t-{\mathcal{T}}+1}_i || {\overline{\mathcal{SE}}}^{t-{\mathcal{T}}+1}_j)}_\text{${\overline{\mathcal{E}}}_{ij}^{t-{\mathcal{T}}+1}$} ||_R \ldots ||_R \underbrace{({\overline{\mathcal{UE}}}^{t-1}_i || {\overline{\mathcal{SE}}}^{t-1}_j)}_\text{${\overline{\mathcal{E}}}_{ij}^{t-1}$} ||_R \underbrace{({\overline{\mathcal{UE}}}^t_i || {\overline{\mathcal{SE}}}^t_j)}_\text{${\overline{\mathcal{E}}}_{ij}^t$})
\end{equation}

\subsubsection{Training of PTE}
We train the PTE with the input embedding ${\mathcal{IE}}_{ij}^t$ and target value $q_{ij}^t$, for all $u_i, s_j$ such that ${\mathcal{Q}}(i,j,t) \neq 0$. 
Fig. \ref{fig:transformer}(b) presents the detail of our transformer encoder network. 
The core component of PTE is multi-head attention (MHA) with $h$ heads as presented in Fig. \ref{fig:transformer}(c).
Each head of MHA comprises scaled dot-product attention (SDPA) \cite{TRANSFORMER}, which is presented in Fig. \ref{fig:transformer}(d). 
The objective of the MHA block is to compute the attention required to give each part of the input across different time-steps. Each head $i$ (where, $i \in \{1, 2, \ldots, h\}$) of MHA computes a tuple comprising query, key, and value, i.e., ($Q^i, K^i, V^i$) as shown below.
\begin{equation}\small
\begin{split}
    (Q^i)_{{\mathcal{T}} \times d_k} = ({\mathcal{IE}}_{ij}^t)_{{\mathcal{T}} \times 4f'} ~(W_Q^i)_{4f' \times d_k}\\
    (K^i)_{{\mathcal{T}} \times d_k} = ({\mathcal{IE}}_{ij}^t)_{{\mathcal{T}} \times 4f'} ~(W_K^i)_{4f' \times d_k}\\
    (V^i)_{{\mathcal{T}} \times d_v} = ({\mathcal{IE}}_{ij}^t)_{{\mathcal{T}} \times 4f'} ~(W_V^i)_{4f' \times d_v}
\end{split}
\end{equation}

\noindent
The tuple ($Q^i, K^i, V^i$) is then forwarded to the SDPA block, which is responsible for computing the attention for the input sequence across $\mathcal{T}$ time-steps. The outcomes of each SDPA block of size ${{\mathcal{T}} \times d_v}$ are now concatenated column-wise across all heads, and fed to the next block to perform a linear operation with weight matrix $W_L$. It may be noted, the MHA block produces an output of size ${\mathcal{T}} \times 4f'$.

The PTE comprises a residual connection that adds the output generated by MHA with ${\mathcal{IE}}_{ij}^t$, which is then passed through a layer normalization (LN) \cite{TRANSFORMER}. The output of this LN is then fed to two 1D convolution layers (Conv1D) connected sequentially. In between these two Conv1D layers, we use a dropout layer for regularization \cite{dropout}. The output of the second Conv1D is added to the output of the previous LN, which is then fed to another LN to generate the final output vector of the PTE.    

In TQP, the input embedding passes through $C_1$ number of PTE blocks sequentially. The output vector of the final PTE block is fed to a global max pooling layer, which is then forwarded to $C_2$ number of fully connected layers to predict the value ${\hat{q}}_{ij}^t$. A detailed configuration of TQP is presented in Section \ref{sec:results}.

\subsection{Outlier Detection and Handling}\label{subsec: outlier_algo}
\noindent
The QoS prediction accuracy is highly sensitive to the outliers present in the dataset \cite{CTF}. In this paper, we employ two different strategies to handle the outliers. To train GCMF and PTE, we use Cauchy loss \cite{CAUCHY-LOSS} to handle outliers, as presented in Eq.s \ref{eq:gcn_loss} and \ref{eq:te_loss}, respectively.

%
{\scriptsize
\begin{equation}\label{eq:gcn_loss}
    {\mathcal{L}}^k_{GCMF} = \sum \limits_{q_{ij}^k \neq 0}\log \left(1+ \left({(q^{k}_{ij} - {\hat{q}}^k_{ij})}/{\gamma_s}\right)^2\right)
\end{equation}
}{\scriptsize
\begin{equation}\label{eq:te_loss}
    {\mathcal{L}}_{PTE} = \sum \limits_{q_{ij}^t \neq 0}\log \left(1+ \left({(q^{t}_{ij} - {\hat{q}}^t_{ij})}/{\gamma_t}\right)^2\right)
\end{equation}}

\noindent 
where, $\gamma_s$ and $\gamma_t$ are hyper-parameters needed to be tuned externally. 

\noindent
Furthermore, we employ isolation forest algorithm \cite{isolation-forest} to detect the outliers present in the dataset for a given outlier ratio $\lambda$, where $\lambda$ is another hyper-parameter used to decide the percentage of outliers required to be removed to evaluate the performance of our framework. In our experiment, we show the performance of TPMCF for different values of $\lambda$. 

In the next section, we present the performance of TPMCF empirically.

\section{Experiments}\label{sec:results}
\noindent
The implementation of our framework was performed using TensorFlow v2.6.2 with Python 3.6.9. For training purposes, NVIDIA's Quadro RTX 3000/PCIe/SSE2 GPU with 1920 cores, and 6 GB memory were used. We evaluate the trained model using i9-10885H @ 2.40 GHz$\times$16 processor with x86\_64 CPU with 128 GB RAM. 

\subsection{Experimental Setup}
\noindent 
We now discuss the employed dataset followed by the performance metric, train-test split-up, and model configuration.

\noindent
{\textbf{Dataset description}}: To validate the effectiveness of our proposed framework, we used the WSDREAM-2 \cite{WSDREAM} dataset, which comprises two QoS attributes, namely, Response Time (RT) and Throughput (TP). For both attributes, the QoS values are recorded for 64 different time-steps, each at a 15-minute interval for globally distributed 142 users and 4500 services. The statistical details of the dataset are shown in Table \ref{tab:dataset}.

\begin{table}[!h]
\centering
\caption{WSDREAM-2 dataset statistics \cite{WSDREAM}}
\begin{tabular}{c|c|c || c|c|c }
\hline 
  & RT & TP &  & RT & TP \\ \hline \hline
 \# User ($n$) & 142 & 142 & Mean & 3.177 & 11.345 \\ 
 \# Service ($m$) & 4500 & 4500 & Median & 0.442 & 1.852 \\ 
 \# Time-step & 64 & 64 & SD & 6.128 & 54.276 \\ 
 Min & 0.001 & 0.000036 & Max & 20.0 & 6726.833\\ \hline
 \multicolumn{6}{r}{\tiny{SD: Standard Deviation}}
\end{tabular}
\label{tab:dataset}
\end{table}

\noindent
\textbf{Performance Metric}: 
Here, we adopted Mean Absolute Error (MAE), a widely used statistical performance metric for QoS prediction, which is defined as follows.

\noindent 
\begin{equation}\scriptsize
    MAE = \left(\sum_{(u_i, s_j, t_k) \in TD}\lvert q_{ij}^k - {\hat{q}}_{ij}^k \rvert\right) / |TD|
\end{equation}

\noindent
where, $TD$ is the test dataset. The smaller value for MAE indicates better accuracy. 


\noindent
\textbf{Train-test split-up}: In our experimentation, we used two different train-test split-ups. We divided each dataset into $10\%:90\%$ and $20\%:80\%$ of training-testing ratios for our experiment. Table \ref{tab:dataset_} presents the details of these datasets.

%
\begin{table}[!h]\scriptsize

    \centering
    \caption{Dataset Description}
    \begin{tabular}{c|c|c || c|c|c}
     \hline
     Name & Parameter & Train : Test & Name & Parameter & Train : Test \\
     \hline\hline
     RT-10 &  RT & $10\%:90\%$ & RT-20 &  RT & $20\%:80\%$ \\ 
     TP-10 &  TP & $10\%:90\%$ & TP-20 &  TP & $20\%:80\%$ \\\hline
    \end{tabular}
    \label{tab:dataset_}
\end{table}
\noindent
Each experiment was performed five times. Finally, the average values were recorded and presented in this paper.

\noindent
\textbf{Configuration of TPMCF}:  We now present the details of various hyper-parameters of TPMCF.
The configuration details for the GCMF and PTE are presented in Table \ref{tab:gcn_parameters}.

\begin{table}[!h] \scriptsize
    \centering
    \caption{Configuration of TPMCF}
    \begin{tabular}{l|l|l|c}
    \hline
       \multicolumn{2}{c|}{\textbf{Parameters}} &  \multicolumn{2}{c}{\textbf{Values}} \\ \hline \hline
       \multicolumn{2}{c|}{QoS Features} & $f_q$ & {100} \\
       \multicolumn{2}{c|}{Collaborative features} & $f_c$ &{50} \\\hline        
        \multirow{4}{*}{\rotatebox{90}{GCMF}} & No. of GConv Units	& \multicolumn{2}{c}{2}\\ \cline{3-4}
        &Dimension of weight matrix & $W^t_1$  & 155 $\times$ 64 \\ 
        &Dimension of weight matrix & $W^t_2$ & 64 $\times$ 64 \\ 
        &Dimension of automated feature vector & $2f'$ & 128\\  \hline
        \multirow{10}{*}{\rotatebox{90}{PTE}} &No. of time-steps & $\mathcal{T}$ & 8 \\ 
        &No. of PTE blocks	& $C_1$ & 4 \\ 
        &No. of heads	& $h$ & 4 \\ 
        &Dimension of query and key 	& $d_k$ & 256 \\ 
        &Dimension of value 	& $d_v$ & 256 \\ \cline{3-4}     
        &\multirow{2}{*}{First Conv 1D} & No. of filters & 4 \\ 
        && Filter size & 3$\times$3 \\ \cline{3-4} 
        &\multirow{2}{*}{Second Conv 1D}
        & No. of filters & $\mathcal{T}$ \\ 
        && Filter size &  1$\times$1 \\ \cline{3-4}         
        &No. of fully connected (FC) layers & $C_2$ & 2 \\ \hline
        \multicolumn{2}{c|}{ Optimizer (GCMF, PTE)} & \multicolumn{2}{c}{AdamW \cite{adamw}} \\ \hline
        
    \end{tabular}
    \label{tab:gcn_parameters}
\end{table}

\noindent
We used $\lambda \text{~(outlier ratio)} = 0.1$ throughout our experiments unless otherwise specified. 
We also present a few experiments changing the values of some hyper-parameters to show their impact on prediction accuracy.

\subsection{Experimental Analysis}
\noindent
We now analyze the performance of TPMCF and demonstrate the comparative study with respect to the major state-of-the-art methods (SoA).

\subsubsection{Performance of TPMCF and Comparison with SoA}
\label{subsec:comparision_SoA}
Here, we present the performance of TPMCF in terms of prediction time and accuracy, and the training time of the models. 
Table \ref{tab:comparisonSOA} presents the comparison between SoA and our framework in terms of prediction accuracy. In this table, we have categorized the SoA methods in terms of their usage of features:  
(a) \emph{QoS\_M:} methods that used only QoS features derived from the QoS invocation log $\mathcal{Q}$, and 
(b) \emph{QoS\_Context\_M:} methods employing the contextual information of users and services along with the QoS features.

{\textbf{Comparison with SoA in terms of prediction accuracy}}: As evident from Table \ref{tab:comparisonSOA}, TPMCF outperformed all the SoA. We observed an improvement of 5.46\%, 16.36\%, 25.55\%, and 34.02\% of TPMCF over the second best SoA with QoS features as reported in Table \ref{tab:comparisonSOA} on datasets RT-10, RT-20, TP-10, and TP-20, respectively. Similarly, we reported an improvement of our framework over the second-best SoA with both the QoS and contextual features in Table \ref{tab:comparisonSOA}. The final row of Table \ref{tab:comparisonSOA} shows the normalized MAE of TPMCF for all four cases of RT and TP datasets, which also shows that our framework is generalized enough.

\begin{figure}[!b]
    \centering
    \includegraphics[width=0.6\linewidth]{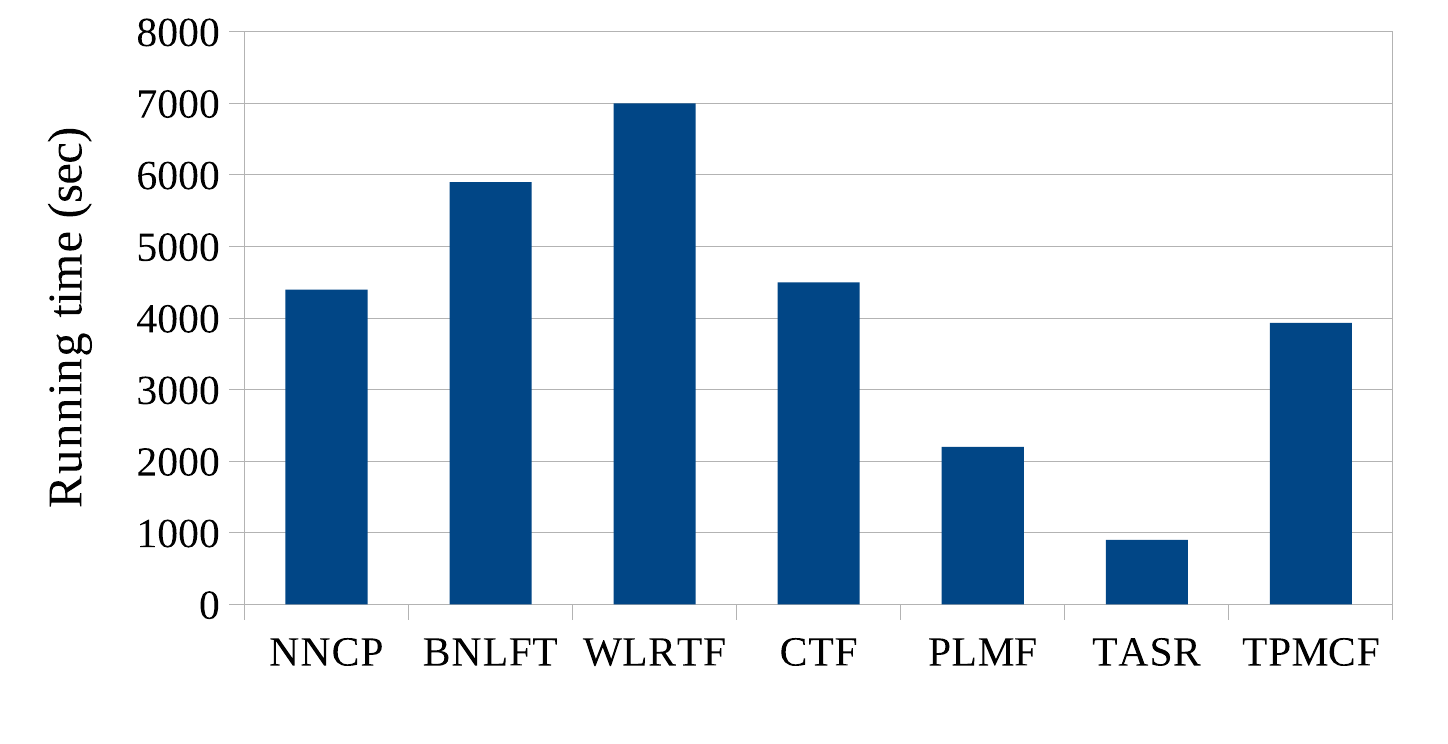}
    \caption{Training time comparison of TPMCF with SoA}
    \label{fig:runtime}
\end{figure}

{\textbf{Comparison with SoA in terms of learning time}}: Fig. \ref{fig:runtime} shows the comparison between SoA and our framework in terms of the training time of the models. 
We have the following observations from Fig. \ref{fig:runtime}. 
    
\emph{(i)} TPMCF had less training time as compared to four SoA methods (i.e., NNCP \cite{NNCP}, BNLFT \cite{BNLFT}, WLRTF \cite{WLRTF}, AND CTF \cite{CTF}). Out of these four methods, NNCP \cite{NNCP} is the fastest in terms of training time. TPMCF achieved 1.12$\times$ speed-up and an average improvement of 62.11\% over all four datasets as compared to NNCP.
    
\emph{(ii)} CTF \cite{CTF}, on the other hand, is the best in terms of prediction accuracy among the four above-mentioned SoA methods. As compared to CTF, TPMCF achieved 1.14$\times$ speed-up and an average improvement of 40.64\% over all four datasets. 
    
\emph{(iii)} TPMCF was slower than PLMF \cite{PLMF} and TASR \cite{TASR} with speed-degradation of 0.6$\times$ and 0.2$\times$, respectively. However, on average, TPMCF achieved an improvement of 33.38\% and 80.78\% over PLMF and TASR, respectively. 
    
{\textbf{Comparison with SoA in terms of prediction time}}: For TPMCF, the training is performed in offline mode. Therefore, the longer training time does not affect the actual performance of the framework. The performance of TPMCF is measured by the prediction time. The average prediction time for TPMCF is $4.1 \times 10^{-4}$ seconds, which is reasonably well as compared to the minimum response time of the service (which is of $10^{-3}$ seconds).    

\begin{table}[!t]\scriptsize
    \centering
    \caption{Performance of TPMCF and Comparison with SoA in terms of prediction accuracy (MAE)}
    \begin{tabular}{c|l|c|c|c|c}
    \hline
    {Category} & {Methods} &  {RT-10} & {RT-20} & {TP-10} & {TP-20} \\
    \hline \hline
    \multirow{12}{*}{\emph{QoS\_M}} 
    & TASR \cite{TASR}   & 2.8188 &	2.7120 & 4.3265 &	3.6419 \\ 
    & WSPred \cite{WSPRED}  & 2.4990 &  2.3000 & 8.0131 & 7.6000 \\ 
    & OPST \cite{OPST}  &	1.1722  &	1.1587  &	3.2762  &	3.1193   \\ 
    & BNLFT \cite{BNLFT} & 1.0828 &		1.0575 &		1.4241 &	1.3935 \\ 
    & NNCP \cite{NNCP}  & 1.0796 &	1.0536 &	2.6401 & 2.5797  \\     
    & WLRTF \cite{WLRTF} & 1.0560 &	1.0437 &	2.9576 &	2.9569 \\ 
    & RNCF \cite{RNCF}   & 1.0100 &		0.9580 &	- &		-  \\ 
    & CTF \cite{CTF} & 0.9215 &	0.8981 & \textcolor{blue}{1.3567} &	\textcolor{blue}{1.1945}   \\ 
    & CARP \cite{carp} & 0.7709 & 0.6992 & - & -  \\ 
    & PLMF \cite{PLMF}  &	0.6786 &		0.6444 &	- &		- \\ 
    & RTF \cite{RTF} & 0.6300  &	0.5350 & 4.0250 &		3.799 \\ 
    & DeepTQSP \cite{DeepTSQP} & {0.5794} & {0.4526} & - & -\\ 
    & TUIPCC \cite{TUIPCC} & 0.5767 & 0.6970  & 3.7573 & 3.7177 \\ 
    & GAT + GRU \cite{icsoc22} & \textcolor{blue}{0.5260} & \textcolor{blue}{0.4620}  & - & - \\ \hline
    \multirow{2}{*}\emph{QoS\_Context\_M}
    & Mul-TSFL \cite{Mul-TSFL}  &	0.8972 &	\textcolor{blue}{0.8097} &	- &		-  \\ 
    & QSPC \cite{QSPC}  & \textcolor{blue}{0.8341}   &	0.8231 &	\textcolor{blue}{3.0604} & \textcolor{blue}{2.9624} \\ \hline

   \multirow{1}{*}{\emph{QoS\_M}} 
   & \textbf{TPMCF}   & \textbf{0.4973} &	\textbf{0.3864} & \textbf{1.0101} &	\textbf{0.7881} \\ \hline 
    
    {\emph{QoS\_M}} 
    & \textbf{Improvement}* &	\textbf{5.46\%} & \textbf{16.36\%} &	\textbf{25.55\%} & \textbf{34.02\%}\\ \hline 
    \emph{QoS\_Context\_M}
    & \textbf{Improvement}* &	\textbf{40.38\%} & \textbf{52.28\%} &	\textbf{66.99\%} & \textbf{73.40\%}\\ \hline
    \multicolumn{2}{c|}{NMAE} & 0.1565 & 0.1216 & 0.0890 & 0.0695 \\ \hline 
    \multicolumn{6}{r}{*\textbf{Improvement} over the  \textcolor{blue}{second-best method}}\\    
    \end{tabular}
    \label{tab:comparisonSOA}
\end{table}

\subsubsection{Outlier Analysis}
\label{subsec:outlier_analysis}
We now present the robustness of TPMCF in the presence of outliers. Table \ref{tab:loss_impact} presents the performance of TPMCF with four different loss functions for various values of the outlier ratio ($\lambda$) on RT-10 dataset. It may be noted that mean-squared-error (MSE) \cite{mse} is outlier sensitive, and hence, the performance of TPMCF with MSE as the loss function was the worst. Mean-absolute-error (MAE), Huber loss \cite{Huber1992}, and Cauchy loss are comparatively outlier resilient. Cauchy loss \cite{CAUCHY-LOSS}, however, performed the best compared to the other loss functions (refer to Table \ref{tab:loss_impact}).

\begin{table}[!h]\scriptsize
\caption{Impact of loss function on RT-10 dataset  (MAE)}
    \centering
    \begin{tabular}{c|c|c|c|c|c|c}
    \hline 
    \multirow{2}{*}{Loss Function}  & \multicolumn{6}{c}{Outlier ratio ($\lambda$)}\\\cline{2-7}
         & 0 & 0.02 & 0.04 & 0.06 & 0.08 & 0.1 \\ \hline\hline
        MSE  & 0.9419 & 0.8443 & 0.7904 & 0.7425 & 0.7063 & 0.6796 \\ 
        MAE &  0.8430 & 0.7421 & 0.6887 & 0.6417 & 0.6057 & 0.5800 \\ 
        Huber & 0.8409 & 0.7384 & 0.6869 & 0.6392 & 0.6018 & 0.5742 \\ 
        Cauchy & 0.8132 & 0.6985 & 0.6279 & 0.5670 & 0.5255 & 0.4973 \\ \hline
    \end{tabular}    
    \label{tab:loss_impact}
\end{table}

\begin{table}[!t]\scriptsize
\caption{Performance (MAE) analysis over different $\lambda$}
    \label{tab:outlier_impact}
    \centering
    \begin{tabular}{c|c|c|c|c|c|c}
    \hline 
       \multirow{2}{*}{Datasets} & \multicolumn{6}{c}{Outlier ratio ($\lambda$)}\\\cline{2-7}
        & 0 & 0.02 & 0.04 & 0.06 & 0.08 & 0.1 \\ \hline\hline
        RT-10  & 0.8132 & 0.6985 & 0.6279 & 0.5670 & 0.5255 & 0.4973  \\ 
        RT-20  & 0.6159 & 0.5326 & 0.4834 & 0.4399 & 0.4072 & 0.3864 \\ 
        TP-10  & 6.4200 & 2.7457 & 1.7873 & 1.3801 & 1.1490 & 1.0101 \\ 
        TP-20  & 5.4620 & 2.3853 & 1.5100 & 1.1254 & 0.9179 & 0.7881 \\ \hline
    \end{tabular}    
\end{table}

\begin{figure}[!t]
    \centering
    (a)\includegraphics[width=0.4\linewidth]{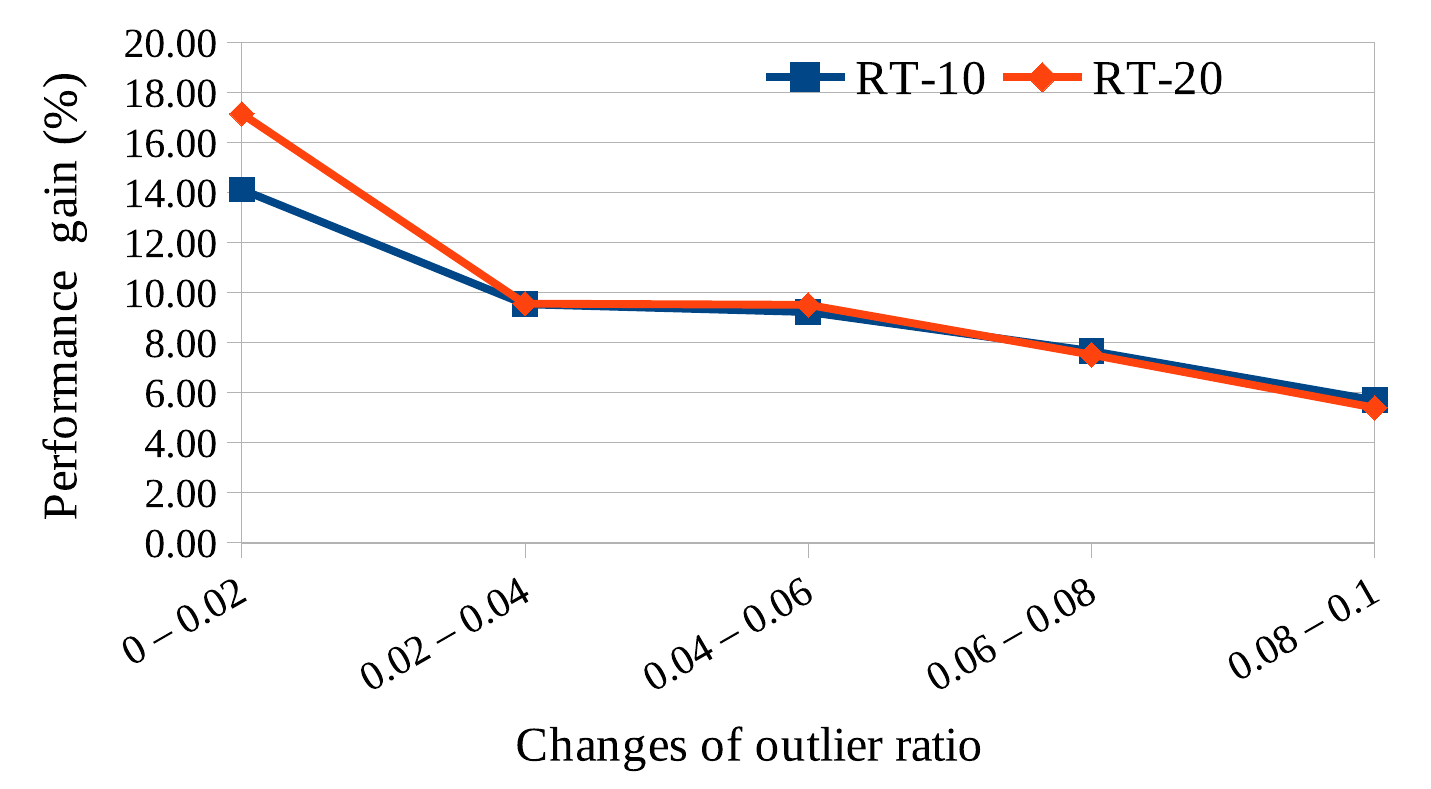}
    (b)\includegraphics[width=0.4\linewidth]{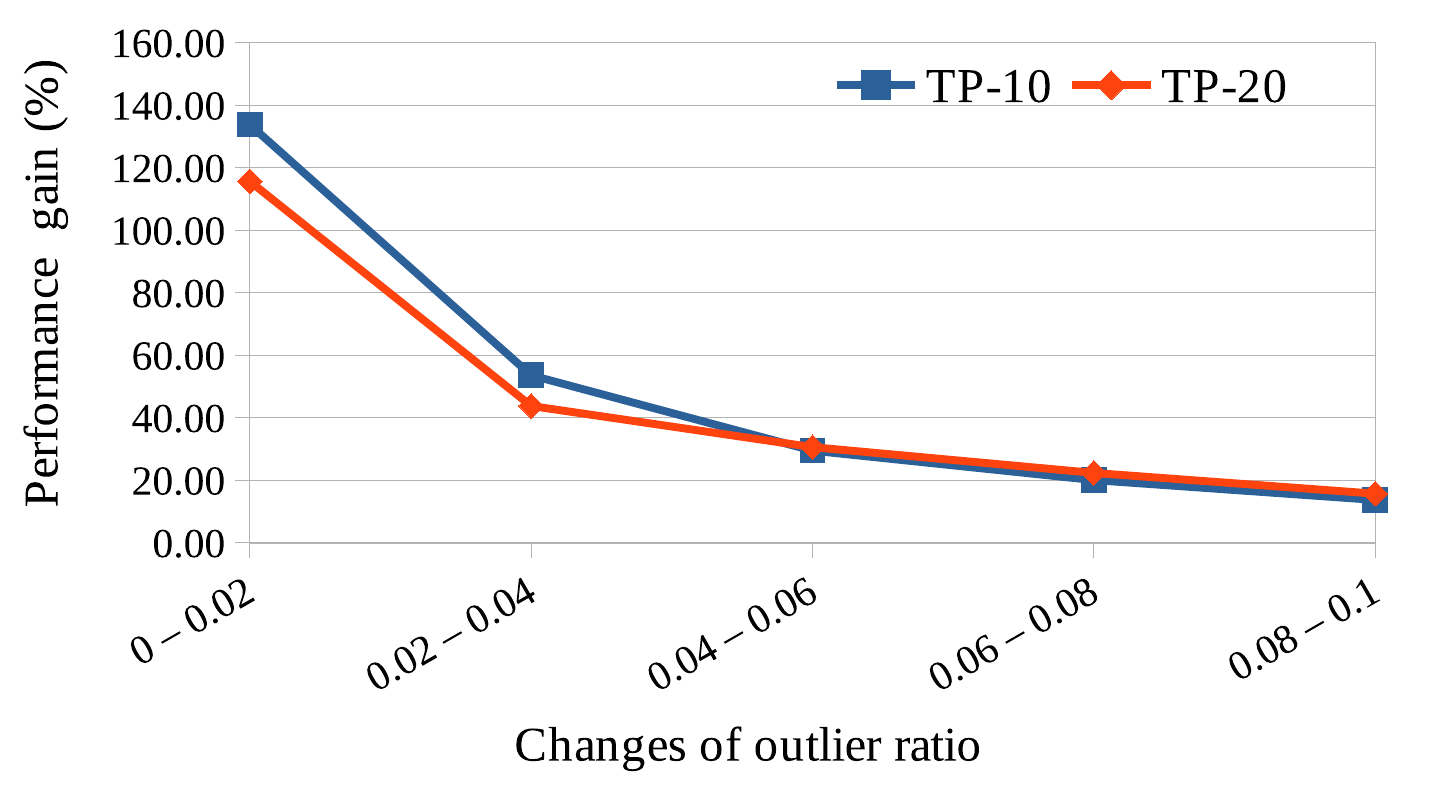}
    \caption{Performance gain with change in outlier ratio $\lambda$ on (a) RT and (b) TP datasets}
    \label{fig:impact_outlier}
\end{figure}

\noindent
Table \ref{tab:outlier_impact} presents the impact of the outliers on the prediction accuracy of TPMCF for all four datasets. As observed from Table \ref{tab:outlier_impact}, the outliers have a severe impact on the performance of TPMCF. In Fig. \ref{fig:impact_outlier}, we report the performance gain with the increasing values of outlier ratio ($\lambda$). 
As observed from Fig. \ref{fig:impact_outlier}, when we removed the first 2\% outliers ($\lambda = 0.02$), we achieved maximum performance gain ($PG$) that is defined as follows:
\begin{equation}
    PG (m_1, m_2) = ((m_1 - m_2)/m_2) \times 100\%
\end{equation}
where, $m_1$ and $m_2$ represent the MAE of TPMCF after removing outliers with the ratio $\lambda_1$ and $\lambda_2$, respectively ($\lambda_1 < \lambda_2$). As we removed more outliers, the performance gain gradually decreased, as evident from decreasing trends of the curves shown in Figures \ref{fig:impact_outlier}(a), (b).
With the change in $\lambda$ from 0.08 to 0.1, we achieved the least performance gain.
In other words, the initial 2\% outliers influenced the performance measure (i.e., MAE value) more than the rest. The rate of change of performance improvement decreased with the increase in the outlier ratio. This explains that the outlier detection algorithm is powerful enough to identify the appropriate set of outliers.

\begin{figure}[!t]
    \centering
    (a) \includegraphics[width=0.4\linewidth]{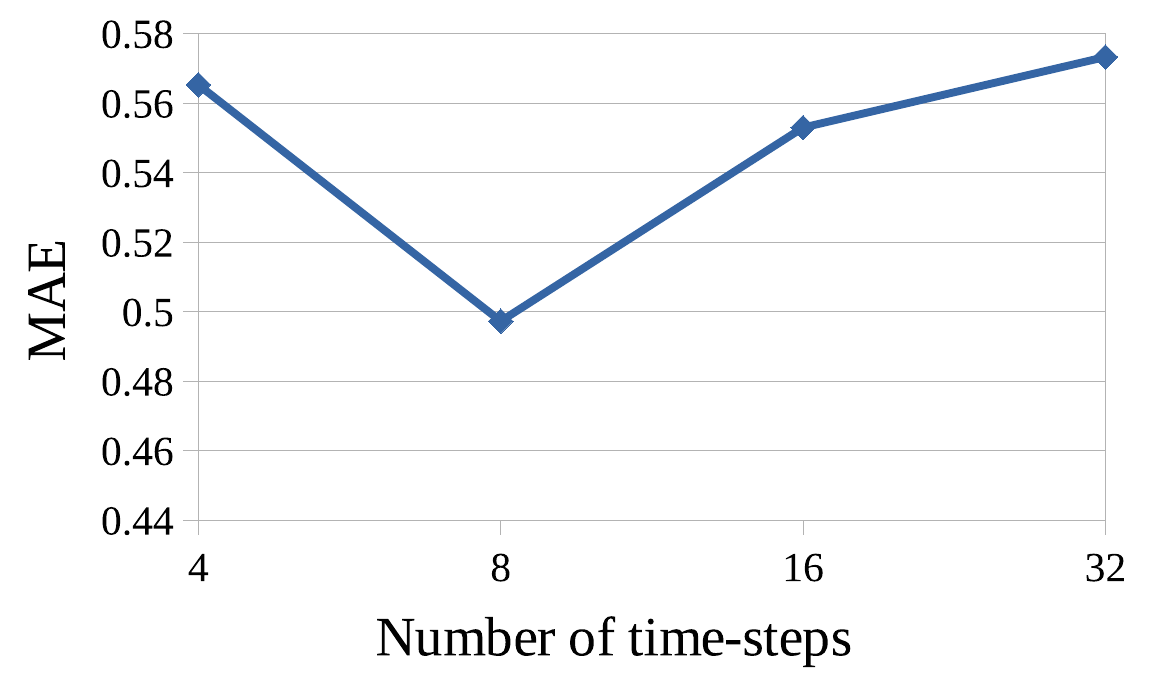}
    (b) \includegraphics[width=0.4\linewidth]{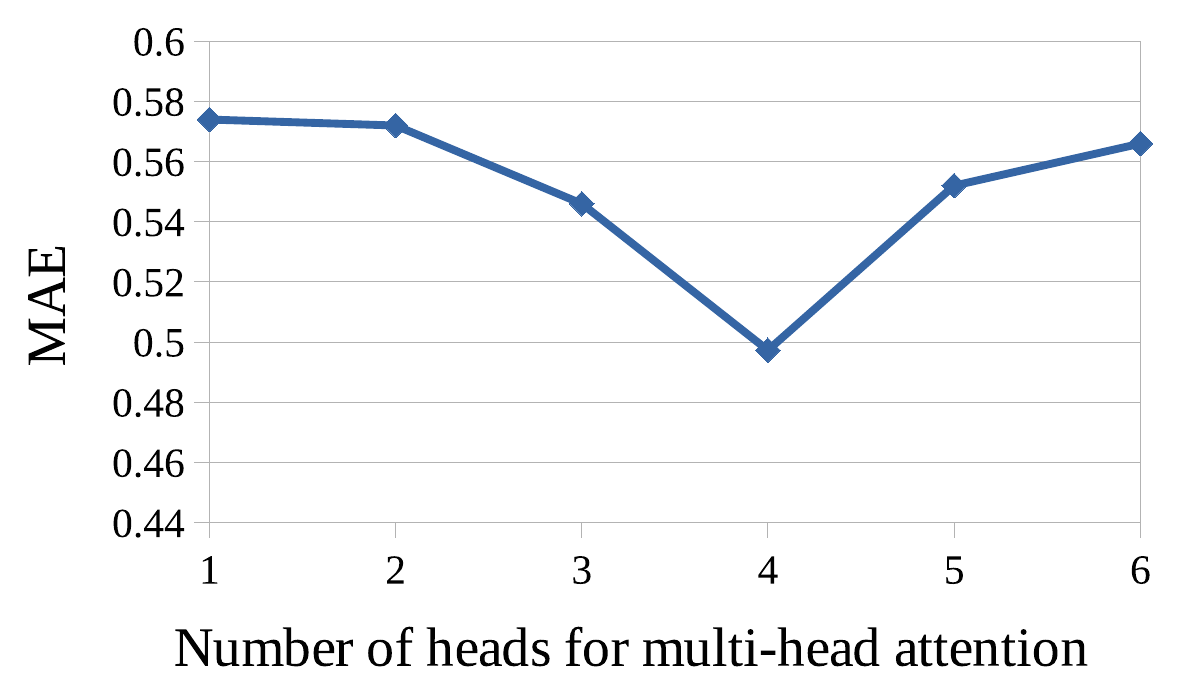}
    \caption{Impact of (a) number of time-steps $\mathcal{T}$, (b) number of heads $h$ in multi-head attention on model performance}
    \label{fig:impact}
\end{figure}

\begin{figure}[!t]
    \centering
    (a) \includegraphics[width=0.4\linewidth]{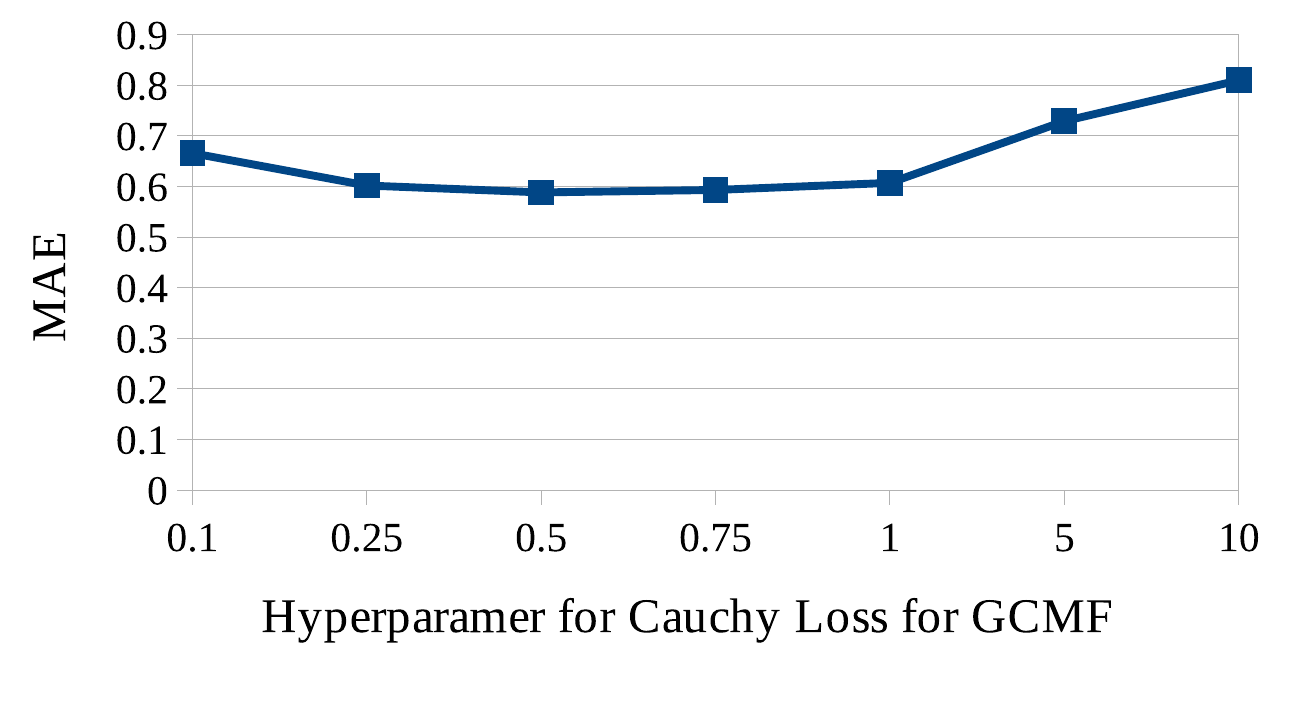}
    (b) \includegraphics[width=0.4\linewidth]{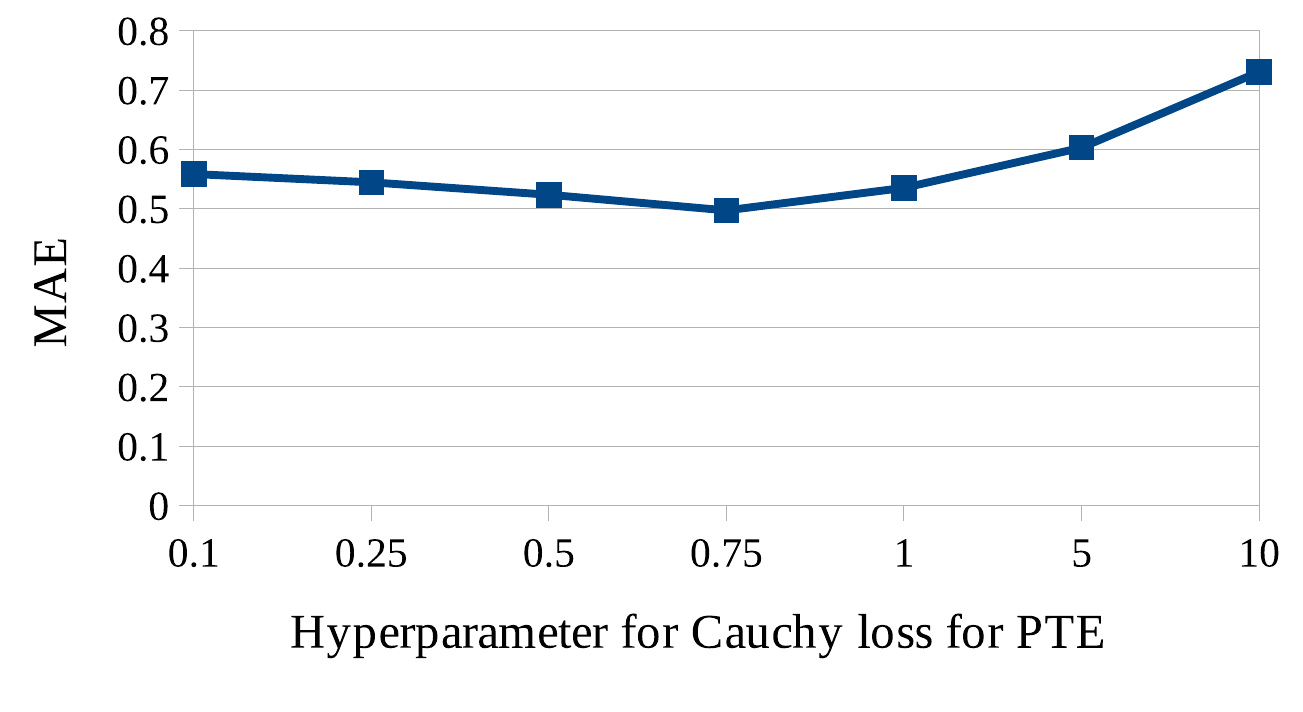}
    \caption{Impact of (a) $\gamma_s$ on performance of GCMF, (b) $\gamma_t$ on performance of TPMCF}
    \label{fig:impact_gamma}
\end{figure}

\subsubsection{Impact of Hyper-parameters}\label{subsec:impact}
This section analyzes the impact of a few hyper-parameters on the performance of TPMCF.  

\noindent
\textbf{Impact of the number of time-steps}: Fig. \ref{fig:impact}(a) shows the MAE obtained by TPMCF on the RT-10 dataset with the change in the number of time-steps ($\mathcal{T}$) considered to construct the input embedding for TQP. We achieved the best performance for ${\mathcal{T}} = 8$. As observed from Fig. \ref{fig:impact}(a), with the increase in the value of $\mathcal{T}$, the MAE was decreasing till 8 time-steps, and then it started increasing for $\mathcal{T} =  16$ and 32.

\noindent
\textbf{Single vs. Multi-Head Attention}: To show the impact of the number of heads ($h$), we varied the value of $h$ from  1 to 6. The results are reported in Fig. \ref{fig:impact}(b) for the RT-10 dataset. 
Initially, with the increase in the value of $h$, the MAE value decreased till $h = 4$, and it started increasing thereafter, as evident from Fig. \ref{fig:impact}(b).

\noindent
\textbf{Impact of $\gamma_s$ and $\gamma_t$}: Fig. \ref{fig:impact_gamma}(a) shows the impact of $\gamma_s$ on the performance of GCMF on the RT-10 dataset. As observed from the figure, GCMF achieved the best prediction accuracy for $\gamma_s = 0.5$. Similarly, Fig. \ref{fig:impact_gamma}(b) shows the impact of $\gamma_t$ on the performance of TPMCF on the RT-10 dataset while keeping $\gamma_s$ as constant. As observed from Fig. \ref{fig:impact_gamma}(b), TPMCF achieved the best prediction accuracy for $\gamma_t = 0.75$.

\subsubsection{Ablation Study}\label{subsec:ablation_study}
We now present the rationale behind our model selection and the justification for having various modules of TPMCF through the ablation study. 

\begin{table}[!h]\scriptsize
    \caption{Study on Model Selection on RT-10 (MAE)}
    \label{tab:model_impact}
    \centering
    \begin{tabular}{c|c|c|c|c}
    \hline 
    GAT+PTE & GConv+PTE & GCMF+LSTM & GCMF+GRU & {{TPMCF}} \\
    \hline\hline
    1.1161 & 0.8092 & 0.6111 & 0.6171 & {{0.4973}} \\\hline
    \end{tabular}
\end{table}

{\textbf{Model Selection}}: Table \ref{tab:model_impact} presents the performance of TPMCF in terms of MAE as compared to other models that can be used to generate spatial and temporal features. We have the following observations from Table \ref{tab:model_impact}.

\emph{(i)} We first compare GCMF with graph attention network (GAT) \cite{gat} and graph convolutional network (GConv) \cite{trqp} on the RT-10 dataset. As evident from Table \ref{tab:model_impact}, TPMCF achieved an improvement of 55.44\% and 38.54\% over GAT+PTE and GConv+PTE architectures, respectively. This experiment shows the effectiveness of GCMF in TPMCF. 

\emph{(ii)} We further implemented the temporal QoS prediction block using LSTM \cite{LSTM} and GRU \cite{GRU}. TPMCF achieved an improvement of 18.62\% and 19.41\% over GCMF+LSTM and GCMF+GRU architectures, respectively. This experiment explains the importance of the PTE as part of the temporal QoS prediction block.

\begin{table}[!h]\scriptsize
    \caption{Module ablation study (MAE)}
    \label{tab:module_impact}
    \centering
    \begin{tabular}{c|c|c|c}
    \hline 
   \multirow{3}{*}{Datasets} & \multicolumn{3}{c}{Features} \\ \cline{2-4}
    & Spatial  & Temporal  & Spatial + Temporal \\ \cline{2-4}
    & GCMF & PTE & TPMCF \\
    \hline\hline
    RT-10 & 0.5884 & 0.5842 & 0.4973 \\ 
    RT-20 & 0.4762 & 0.4796 & 0.3864 \\  
    TP-10 & 1.1616 & 1.1879 & 1.0101 \\  
    TP-20 & 0.8877 & 0.8506 & 0.7881 \\ \hline 
    \end{tabular}
\end{table}

{\textbf{Module Ablation}}:
Table \ref{tab:module_impact} presents the ablation study for various modules in TPMCF. We have the following observations from Table \ref{tab:module_impact}.
    
\emph{(i)} TPMCF achieved on average 14.65\% improvement over the stand-alone GCMF, which justifies the requirement of temporal QoS prediction block.
    
\emph{(ii)} TPMCF achieved on average 14.16\% improvement over the stand-alone transformer encoder with temporal features. This implies the necessity of capturing the spatial features along with the temporal dependencies. 
    
 


{\textbf{Feature Ablation}}:
Table \ref{tab:feature_impact} presents the feature ablation study for TPMCF. As observed from Table \ref{tab:feature_impact}, TPMCF with the combined features outperformed the same model with individual feature categories, as discussed in Section \ref{subsubsec:features}. On the one hand, this experiment shows the significance of using the collaborative features derived from domain knowledge apart from the other auto-extracted features used for QoS prediction. On the other hand, it justifies using the different sets of features extracted to employ in TPMCF.

\begin{table}[!h]\scriptsize
    \caption{Feature ablation study (MAE)}
    \label{tab:feature_impact}
    \centering
    \begin{tabular}{c|c|c|c|c}
    \hline 
    \multirow{2}{*}{Datasets} & Statistical  & QoS  & Collaborative Features & Combined  \\
    & Features & Features & using Correlations & Features \\
    \hline\hline
    RT-10 & 0.5939 & 0.5821 & 0.5771 & 0.4973 \\
    RT-20 & 0.4517 & 0.4093 & 0.4104 & 0.3864  \\  
    TP-10 & 1.4249 & 1.1350 & 1.0891 & 1.0101  \\ 
    TP-20 & 1.1112 & 0.8281 & 0.8590 & 0.7881  \\ \hline 
    \end{tabular}
\end{table}

\noindent
In summary, TPMCF achieved higher prediction accuracy than the SoA methods, all the while maintaining reasonably fast prediction times.

\section{Related Works}\label{sec:related_work}
\noindent
In this section, we present a brief literature survey on QoS prediction. We broadly classify the QoS prediction into two categories: static and dynamic. We illustrate each of these categories below.

\subsection{Static QoS Prediction}\noindent
In static QoS prediction, methods consider the QoS parameter of service varies across only users \cite{hsanet,tan} and propose effective frameworks for prediction \cite{DAFR-2020,CluCF,OFFDQ}. 

However, it is important to note that the QoS parameter for a user-service pair rarely remains constant over time. It can fluctuate due to several factors, such as service status (e.g., workload and the number of users accessing the service simultaneously) and network congestion \cite{WSPRED,CTF}. As a result, these static methods prove inadequate for predicting QoS values in dynamic QoS environments, where the QoS parameters change over time.

\subsection{Dynamic/ Temporal QoS Prediction}\noindent
Dynamic QoS prediction, often referred to as temporal QoS prediction, takes into account the temporal dependencies among QoS data. Here, we will introduce several SoA strategies for addressing temporal QoS-related challenges.

\noindent
\subsubsection{Temporal Smoothing (TS)-based QoS Prediction}
TS-based methods \cite{TUIPCC,TF-KMP}, commonly rely on similarity-based Collaborative Filtering (CF) to impute missing values in the temporal QoS invocation log. They subsequently calculate a weighted average of QoS data across previous time-steps. In contrast, TMF, as presented in \cite{TMF}, incorporates user-service latent factors into the data imputation process, diverging from the use of similarity-based CF. However, TS-based methods often struggle to accurately capture the intricate features within temporal QoS invocation sequences, posing a challenge in achieving the desired prediction accuracy.

\noindent
\subsubsection{ARIMA-based QoS Prediction}
TASR \cite{TASR} proposed a time-aware and sparsity-tolerant approach to capture the temporal features using a traditional ARIMA-based time-series forecasting model. However, ARIMA-based models are computationally expensive, assume time series data as stationary, and suffer from low accuracy in long-term forecasts involving a large number of time-steps. 

\noindent
\subsubsection{Tensor Factorization (TF)-based QoS Prediction}
TF-based methods \cite{WSPRED, BNLFT, NNCP} employ low-rank tensor factorization to derive latent features for users, services, and time, which are subsequently used for predicting QoS values. Ye et al. \cite{CTF} proposed an improved version of TF-based methods by utilizing Cauchy loss \cite{CAUCHY-LOSS} function, known for its robustness against outliers. However, in general, TF-based methods tend to experience reduced prediction accuracy due to their challenges in capturing effective representations of implicit features.

\noindent
\subsubsection{Deep Architecture (DA)-based QoS Prediction}
Recurrent neural network-based architectures excel at extracting temporal features in a sequence-to-sequence manner. Deep architectures for sequence modeling, such as LSTM or GRU, are predominantly employed for time-aware QoS prediction \cite{RTF, Mul-TSFL,QSPC,DeepTSQP,PLMF,OPST,RNCF,SCATSF}. For instance, Mul-TSFL \cite{Mul-TSFL} uses contextual features, such as geolocation, autonomous system number, and country information of users and services, to fill in missing values and applies a multivariate LSTM model to capture temporal features. DeepTSQP \cite{DeepTSQP} leverages binary invocation and similarity features in conjunction with GRU to capture temporal features for QoS prediction. PLMF \cite{PLMF} introduces an online temporal QoS prediction framework that utilizes the LSTM model to predict the QoS.

In general, models based on LSTM and GRU tend to outperform methods like TS, ARIMA, and TF-based approaches. However, deep learning-based models still face challenges in achieving high prediction accuracy due to their reliance on explicit features and limitations in capturing complex relationships within the data. Moreover, they are often associated with extended convergence times, reduced learning efficiency, and higher memory requirements.

Now, let us discuss the various features used in the literature for temporal QoS prediction and their impact and limitations.

\subsection{Various Feature Representation for QoS Prediction}\noindent
The features used in the literature for temporal QoS prediction can be broadly categorized into two groups: explicit features extracted using domain knowledge and auto-extracted features. The explicit features can further be divided into two subcategories: QoS and contextual features. Similarly, the auto-extracted features can be categorized as spatial or temporal. We now discuss each of these categories in more detail.

\subsubsection{QoS Features}
QoS features are typically extracted from the QoS invocation log and are commonly utilized for QoS prediction. These features are known to be effective for this purpose. However, QoS features can sometimes lack sufficient information due to the sparsity of the QoS invocation log. In sparse environments, QoS features derived from matrix factorization tend to be more expressive \cite{OPST,BNLFT}. Many studies in the literature \cite{trqp,RNCF,PLMF,Mul-TSFL,TF-KMP,TUIPCC,DeepTSQP} have leveraged various QoS features for temporal QoS prediction.

Nonetheless, there are certain limitations to using QoS features alone. They often lack collaborative information from other users or services, which can be crucial for accurate predictions. Additionally, QoS features generally do not contain temporal information, making it challenging to achieve high prediction accuracy solely relying on them.

\subsubsection{Contextual Features}
Some methods in the literature  \cite{Mul-TSFL,QSPC,SCATSF} have incorporated contextual features obtained from contextual information related to users and services, such as latitude, longitude, country, autonomous system, and IP addresses of users and services. While contextual information can provide additional valuable insights about users and services, there is a challenge regarding the availability of such contextual information for all users and services. This limitation can restrict the broader applicability of contextual features for temporal QoS prediction.

\subsubsection{Spatial Features}
Spatial features capture information related to neighborhood relationships, which can be valuable for improving prediction accuracy. Some recent papers have made use of spatial features to enhance predictions by exploring these neighborhood relationships through graph convolution. For example, TAN \cite{tan} utilized network topology information among users and services, incorporating IPs and ASs, and applied graph convolution to capture cross-correlations among them, using BiLSTM for QoS prediction. However, this approach had to exclude a significant number of users and services from the experiment due to the lack of contextual information like IPs and ASs.
HSA Net \cite{hsanet} introduced a privacy-preserving QoS prediction method that learned spatial features using a convolutional neural network alongside known user and service information, incorporating hidden state features obtained through Latent Dirichlet Allocation. TRQP \cite{trqp} employed a graph convolution network to extract spatial features for static QoS prediction. However, these approaches did not capture any temporal dependencies, as their primary focus was on solving the static QoS prediction problem. Therefore, these methods are not recommended for addressing dynamic prediction problems.

\subsubsection{Temporal Features}
Several recent methods have incorporated temporal features for dynamic QoS prediction. For instance, GRU was utilized in \cite{DeepTSQP,RTF,RNCF} to learn user-service invocation features, while LSTM was used in \cite{PLMF,OPST} for QoS prediction. 
In \cite{Temporal_Transformer}, transformers were leveraged for QoS prediction in various IoT applications.
However, these methods did not fully explore the collaborative relationships between users and services, which suggests potential room for improvement in prediction accuracy.

In addition to temporal features, some methods have integrated contextual information (e.g., location, IP, ID) \cite{QSPC,Mul-TSFL,STCA} of the target user or service into their predictions. However, as mentioned earlier, the availability of contextual information can be limited, making it challenging to apply it to all user-service pairs in the QoS prediction problem.
 
\subsubsection{Spatio-temporal Features}
A recent paper \cite{icsoc22} delved into incorporating spatio-temporal features for QoS prediction. Hu et al. \cite{icsoc22} leveraged a graph attention network for spatial features alongside a GRU for temporal dependencies. 
While their approach was promising, it is worth noting that adding multi-source collaborative features derived from domain knowledge could potentially enhance prediction accuracy. However, this aspect remained unexplored in contemporary methods. 



\subsection{Positioning of Our Framework}\noindent
In contrast to previous SoA methods, TPMCF stands out by comprehensively leveraging collaborative features among users and services. TPMCF achieves this by combining explicit features derived from domain knowledge with auto-extracted spatio-temporal features. This dual-feature approach enables TPMCF to capture complex, higher-order relationships among QoS data, resulting in high prediction accuracy.

One of the unique aspects of TPMCF is the use of a predictive transformer encoder (PTE) that pays attention to user-service interactions over time. Additionally, TPMCF's incorporation of graph convolution matrix factorization (GCMF) to extract spatial features contributes to improved prediction accuracy. 
The utilization of Cauchy loss as the objective function ensures that TPMCF is resilient to outliers. Furthermore, offline training of GCMF and the temporal QoS prediction module has no impact on prediction time, ensuring faster responsiveness. Lastly, the latent feature representation obtained from autoencoders for training the GCMF module enhances scalability.

\section{Conclusion}\label{sec:conclusion}
\noindent
This paper presents an efficient solution for temporal QoS prediction for service recommendation. 
Specifically, our framework TPMCF leverages multi-source collaborative features comprising explicit features derived using domain knowledge and auto-extracted spatio-temporal features for capturing the triadic relationships among the users, services and time-steps. TPMCF includes a collaborative spatial feature-extraction module (CSFE) and a temporal QoS prediction module (TQP). CSFE is responsible for spatial feature extraction using graph convolutional matrix factorization (GCMF). On the other hand, TQP is accountable for temporal feature extraction using a predictive transformer encoder (PTE), followed by QoS prediction using a fully connected neural network. 
Additionally, TPMCF is competent in addressing a few fundamental challenges in temporal QoS prediction. For example, to deal with the outliers present in the datasets, TPMCF employs Cauchy loss to train the GCMF and PTE. TPMCF is sparsity-tolerant without using any additional data imputation method due to GCMF. The offline training of GCMF and PTE ensures the high responsiveness of TPMCF. This makes TPMCF a better fit for real-time applications. Moreover, the feature dimensionality reduction with the help of autoencoders makes TPMCF highly scalable.
Our extensive experiments on WSDREAM-2 datasets show that we achieved higher prediction accuracy than the major state-of-the-art methods while having reasonably faster training time as compared to other contemporary methods, and negligible prediction time with respect to the response time of the services. 

Although TPMCF achieved high prediction accuracy, there is still scope for improvement. There are a few challenges, that are unaddressed in our paper, for example, the analysis of the trustworthiness of users and services to exploit collaborative relationships. As our future endeavor, we aim to propose a trust-aware temporal QoS prediction framework. 



\bibliographystyle{IEEEtran}
\bibliography{bibliography}

\begin{thebibliography}{10}
\providecommand{\url}[1]{#1}
\csname url@rmstyle\endcsname
\providecommand{\newblock}{\relax}
\providecommand{\bibinfo}[2]{#2}
\providecommand\BIBentrySTDinterwordspacing{\spaceskip=0pt\relax}
\providecommand\BIBentryALTinterwordstretchfactor{4}
\providecommand\BIBentryALTinterwordspacing{\spaceskip=\fontdimen2\font plus
\BIBentryALTinterwordstretchfactor\fontdimen3\font minus
  \fontdimen4\font\relax}
\providecommand\BIBforeignlanguage[2]{{%
\expandafter\ifx\csname l@#1\endcsname\relax
\typeout{** WARNING: IEEEtran.bst: No hyphenation pattern has been}%
\typeout{** loaded for the language `#1'. Using the pattern for}%
\typeout{** the default language instead.}%
\else
\language=\csname l@#1\endcsname
\fi
#2}}

\bibitem{soa}
T.~Erl, \emph{{Service-oriented architecture: a field guide to integrating XML
  and web services}}.\hskip 1em plus 0.5em minus 0.4em\relax Prentice Hall PTR,
  2004.

\bibitem{Survey1-TSC_Zheng}
Z.~Zheng \emph{et~al.}, ``{Web Service QoS Prediction via Collaborative
  Filtering: A Survey},'' \emph{IEEE TSC}, vol.~15, no.~4, pp. 2455--2472,
  2022.

\bibitem{OFFDQ}
S.~Chattopadhyay \emph{et~al.}, ``{OffDQ: An Offline Deep Learning Framework
  for QoS Prediction},'' in \emph{The ACM Web Conference 2022}, ser. WWW
  '22.\hskip 1em plus 0.5em minus 0.4em\relax ACM, 2022, p. 1987–1996.

\bibitem{DAFR-2020}
Y.~Yin \emph{et~al.}, ``{QoS Prediction for Service Recommendation With
  Features Learning in Mobile Edge Computing Environment},'' \emph{IEEE TCCN},
  vol.~6, no.~4, pp. 1136--1145, 2020.

\bibitem{Survey2-TSC}
S.~H. Ghafouri \emph{et~al.}, ``{A Survey on Web Service QoS Prediction
  Methods},'' \emph{IEEE TSC}, vol.~15, no.~4, pp. 2439--2454, 2022.

\bibitem{UPCC}
J.~S. Breese \emph{et~al.}, ``{Empirical Analysis of Predictive Algorithms for
  Collaborative Filtering},'' in \emph{UAI}, 1998, p. 43–52.

\bibitem{IPCC}
B.~Sarwar \emph{et~al.}, ``{Item-Based Collaborative Filtering Recommendation
  Algorithms},'' in \emph{ACM WWW}, 2001, p. 285–295.

\bibitem{WSRec}
Z.~Zheng \emph{et~al.}, ``{QoS-Aware Web Service Recommendation by
  Collaborative Filtering},'' \emph{IEEE TSC}, vol.~4, no.~2, pp. 140--152,
  2011.

\bibitem{NMF}
D.~D. Lee \emph{et~al.}, ``{Learning the parts of objects by non-negative
  matrix factorization},'' \emph{Nature}, vol. 401, pp. 788--791, 1999.

\bibitem{NIMF}
Z.~Zheng \emph{et~al.}, ``{Collaborative Web Service QoS Prediction via
  Neighborhood Integrated Matrix Factorization},'' \emph{IEEE TSC}, vol.~6,
  no.~3, pp. 289--299, 2013.

\bibitem{EFM}
Y.~Wu \emph{et~al.}, ``{An Embedding Based Factorization Machine Approach for
  Web Service QoS Prediction},'' in \emph{Service-Oriented Computing}.\hskip
  1em plus 0.5em minus 0.4em\relax Cham: Springer, 2017, pp. 272--286.

\bibitem{CAHPHF}
R.~R. Chowdhury \emph{et~al.}, ``{CAHPHF: Context-Aware Hierarchical QoS
  Prediction With Hybrid Filtering},'' \emph{IEEE TSC}, vol.~15, no.~4, pp.
  2232--2247, 2022.

\bibitem{TUIPCC}
E.~Tong \emph{et~al.}, ``{A Missing QoS Prediction Approach via Time-aware
  Collaborative Filtering},'' \emph{IEEE TSC}, pp. 1--1, 2021.

\bibitem{TF-KMP}
C.~Wu \emph{et~al.}, ``{Time-Aware and Sparsity-Tolerant QoS Prediction Based
  on Collaborative Filtering},'' in \emph{2016 IEEE ICWS}, 2016, pp. 637--640.

\bibitem{TMF}
S.~Li \emph{et~al.}, ``{Time-Aware QoS Prediction for Cloud Service
  Recommendation Based on Matrix Factorization},'' \emph{IEEE Access}, vol.~6,
  pp. 77\,716--77\,724, 2018.

\bibitem{TASR}
S.~Ding \emph{et~al.}, ``{Time-Aware Cloud Service Recommendation Using
  Similarity-Enhanced Collaborative Filtering and ARIMA Model},'' \emph{Decis.
  Support Syst.}, vol. 107, no.~C, p. 103–115, 2018.

\bibitem{WSPRED}
Y.~Zhang \emph{et~al.}, ``{WSPred: A Time-Aware Personalized QoS Prediction
  Framework for Web Services},'' in \emph{IEEE ISSRE}, 2011, pp. 210--219.

\bibitem{CTF}
F.~Ye \emph{et~al.}, ``{Outlier-Resilient Web Service QoS Prediction},'' in
  \emph{The Web Conf.}, 2021, p. 3099–3110.

\bibitem{BNLFT}
X.~Luo \emph{et~al.}, ``{Temporal Pattern-Aware QoS Prediction via Biased
  Non-Negative Latent Factorization of Tensors},'' \emph{IEEE Trans. on
  Cybernetics}, vol.~50, no.~5, pp. 1798--1809, 2020.

\bibitem{NNCP}
W.~Zhang \emph{et~al.}, ``{Temporal QoS-Aware Web Service Recommendation via
  Non-Negative Tensor Factorization},'' in \emph{ACM WWW}, 2014, p. 585–596.

\bibitem{WLRTF}
X.~Chen \emph{et~al.}, ``{Robust Tensor Factorization with Unknown Noise},'' in
  \emph{IEEE CVPR}, 2016, pp. 5213--5221.

\bibitem{PLMF}
R.~Xiong \emph{et~al.}, ``{Personalized LSTM Based Matrix Factorization for
  Online QoS Prediction},'' in \emph{IEEE ICWS}, 2018, pp. 34--41.

\bibitem{RTF}
Y.~Zhang \emph{et~al.}, ``{Recurrent Tensor Factorization for Time-Aware
  Service Recommendation},'' \emph{Appl. Soft Comput.}, vol.~85, no.~C, 2019.

\bibitem{OPST}
G.~White \emph{et~al.}, ``{Autoencoders for QoS Prediction at the Edge},'' in
  \emph{IEEE PerCom}, 2019, pp. 1--9.

\bibitem{Mul-TSFL}
M.~Wang \emph{et~al.}, ``{A Location-Based Approach for Web Service QoS
  Prediction via Multivariate Time Series Forecast},'' in \emph{IEEE ICSESS},
  2020, pp. 36--39.

\bibitem{QSPC}
B.~Li \emph{et~al.}, ``{QoS Prediction Based on Temporal Information and
  Request Context},'' \emph{Serv. Oriented Comput. Appl.}, vol.~15, no.~3, p.
  231–244, 2021.

\bibitem{DeepTSQP}
G.~Zou \emph{et~al.}, ``{DeepTSQP: Temporal-Aware Service QoS Prediction via
  Deep Neural Network and Feature Integration},'' \emph{Know.-Based Syst.},
  vol. 241, no.~C, 2022.

\bibitem{RNCF}
T.~Liang \emph{et~al.}, ``{Recurrent Neural Network Based Collaborative
  Filtering for QoS Prediction in IoV},'' \emph{IEEE TITS}, vol.~23, no.~3, pp.
  2400--2410, 2022.

\bibitem{GCN}
T.~N. Kipf \emph{et~al.}, ``{Semi-Supervised Classification with Graph
  Convolutional Networks},'' in \emph{ICLR}.\hskip 1em plus 0.5em minus
  0.4em\relax arXiv:1609.02907, 2017.

\bibitem{TRANSFORMER}
A.~Vaswani \emph{et~al.}, ``{Attention is All You Need},'' in \emph{NIPS},
  2017, p. 6000–6010.

\bibitem{WSDREAM}
Z.~Zheng \emph{et~al.}, ``{Investigating QoS of Real-World Web Services},''
  \emph{IEEE TSC}, vol.~7, no.~1, pp. 32--39, 2014.

\bibitem{COSINE-SIMILARITY}
H.~V. Nguyen \emph{et~al.}, ``{Cosine similarity metric learning for face
  verification},'' in \emph{ACCV}.\hskip 1em plus 0.5em minus 0.4em\relax
  Springer, 2010, pp. 709--720.

\bibitem{STACKED-AUTOENCODER}
P.~Vincent \emph{et~al.}, ``{Stacked Denoising Autoencoders: Learning Useful
  Representations in a Deep Network with a Local Denoising Criterion},''
  \emph{J. Mach. Learn. Res.}, vol.~11, p. 3371–3408, 2010.

\bibitem{GRL-BOOK}
W.~L. Hamilton, ``{Graph representation learning},'' \emph{Synthesis Lectures
  on AIML}, vol.~14, no.~3, pp. 1--159, 2020.

\bibitem{dropout}
N.~Srivastava \emph{et~al.}, ``Dropout: a simple way to prevent neural networks
  from overfitting,'' \emph{J. Mach. Learn. Res.}, vol.~15, no.~1, pp.
  1929--1958, 2014.

\bibitem{CAUCHY-LOSS}
X.~Li \emph{et~al.}, ``{Robust Subspace Clustering by Cauchy Loss Function},''
  \emph{TNNLS}, vol.~30, no.~7, pp. 2067--2078, 2019.

\bibitem{isolation-forest}
F.~T. Liu \emph{et~al.}, ``{Isolation-Based Anomaly Detection},'' \emph{ACM
  TKDD}, vol.~6, no.~1, 2012.

\bibitem{adamw}
I.~Loshchilov \emph{et~al.}, ``{Decoupled Weight Decay Regularization},'' in
  \emph{ICLR}.\hskip 1em plus 0.5em minus 0.4em\relax arXiv:1711.05101, 2019.

\bibitem{carp}
J.~Zhu \emph{et~al.}, ``{CARP:} context-aware reliability prediction of
  black-box web services,'' in \emph{IEEE ICWS}.\hskip 1em plus 0.5em minus
  0.4em\relax {IEEE}, 2017, pp. 17--24.

\bibitem{icsoc22}
S.~Hu \emph{et~al.}, ``{Temporal-Aware QoS Prediction via Dynamic Graph Neural
  Collaborative Learning},'' in \emph{ICSOC}, vol. 13740, 2022, pp. 125--133.

\bibitem{mse}
Z.~Wang \emph{et~al.}, ``{Mean squared error: Love it or leave it? A new look
  at signal fidelity measures},'' \emph{IEEE signal processing magazine},
  vol.~26, no.~1, pp. 98--117, 2009.

\bibitem{Huber1992}
P.~J. Huber, \emph{{Robust Estimation of a Location Parameter}}.\hskip 1em plus
  0.5em minus 0.4em\relax Springer New York, 1992, pp. 492--518.

\bibitem{gat}
P.~Velickovic \emph{et~al.}, ``{Graph Attention Networks},'' in
  \emph{ICLR}.\hskip 1em plus 0.5em minus 0.4em\relax arXiv:1710.10903, 2018.

\bibitem{trqp}
S.~Kumar \emph{et~al.}, ``{TRQP: Trust-Aware Real-Time QoS Prediction Framework
  Using Graph-Based Learning},'' in \emph{ICSOC}, vol. 13740, 2022, pp.
  143--152.

\bibitem{LSTM}
S.~Hochreiter \emph{et~al.}, ``{Long Short-Term Memory},'' \emph{Neural
  Comput.}, vol.~9, no.~8, p. 1735–1780, 1997.

\bibitem{GRU}
J.~Chung \emph{et~al.}, ``{Empirical Evaluation of Gated Recurrent Neural
  Networks on Sequence Modeling},'' \emph{arXiv:1412.3555}, 2014.

\bibitem{hsanet}
Z.~Wang \emph{et~al.}, ``{HSA-Net: Hidden-State-Aware Networks for
  High-Precision QoS Prediction},'' \emph{{IEEE} TPDS}, vol.~33, no.~6, pp.
  1421--1435, 2022.

\bibitem{tan}
J.~Li \emph{et~al.}, ``{Topology-Aware Neural Model for Highly Accurate QoS
  Prediction},'' \emph{{IEEE} TPDS}, vol.~33, no.~7, pp. 1538--1552, 2022.

\bibitem{CluCF}
C.~Yu \emph{et~al.}, ``{CluCF: A Clustering CF Algorithm to Address Data
  Sparsity Problem},'' \emph{Serv. Oriented Comput. Appl.}, vol.~11, no.~1, p.
  33–45, 2017.

\bibitem{SCATSF}
J.~Zhou \emph{et~al.}, ``{Spatial Context-Aware Time-Series Forecasting for QoS
  Prediction},'' \emph{{IEEE} TNSM}, vol.~20, no.~2, pp. 918--931, 2023.

\bibitem{Temporal_Transformer}
A.~Hameed \emph{et~al.}, ``{Toward QoS Prediction Based on Temporal
  Transformers for IoT Applications},'' \emph{IEEE TNSM}, vol.~19, no.~4, pp.
  4010--4027, 2022.

\bibitem{STCA}
Q.~Zhou \emph{et~al.}, ``{Spatio-temporal context-aware collaborative QoS
  prediction},'' \emph{Future Gener. Comput. Syst.}, vol. 100, pp. 46--57,
  2019.

\end{thebibliography}
\vspace{-0.5in}
\begin{IEEEbiography}[{\includegraphics[width=0.8in,height=0.8in,clip,keepaspectratio]{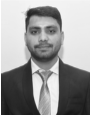}}]
{Suraj Kumar} is pursuing his Ph.D. in CSE from the IIT Indore, India. He received his B.Tech and M.Tech, both in CSE, from Aligarh Muslim University, India, in 2017 and 2019, respectively. His research interests include Services Computing, Machine Learning, and Graph Representation Learning.
\end{IEEEbiography}
\vspace{-0.5in}
\begin{IEEEbiography}[{\includegraphics[width=0.8in,height=0.8in,clip,keepaspectratio]{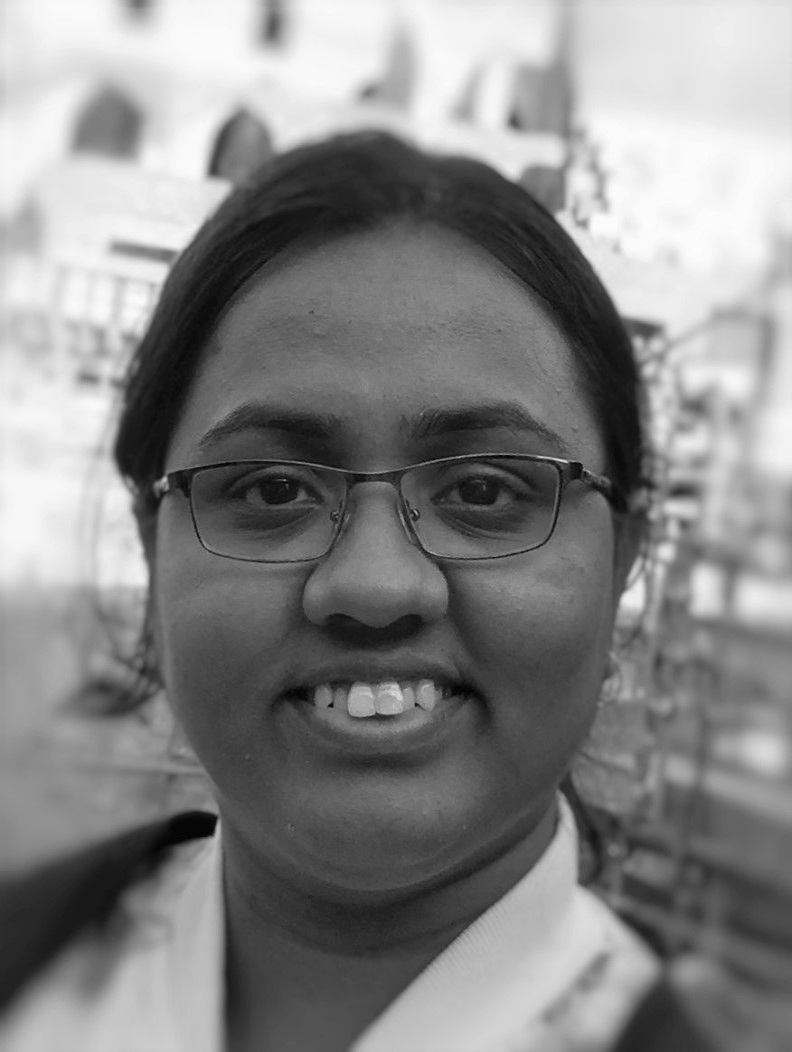}}]
{Soumi Chattopadhyay} (Member, IEEE) received her Ph.D. from the Indian Statistical Institute in 2019. Currently, she is an assistant professor at IIT Indore, India. Her research interests include Services Computing, Artificial Intelligence, Machine Learning, and Deep Learning.
\end{IEEEbiography}

\vspace{-0.5in}
\begin{IEEEbiography}[{\includegraphics[width=0.8in,height=0.8in,clip,keepaspectratio]{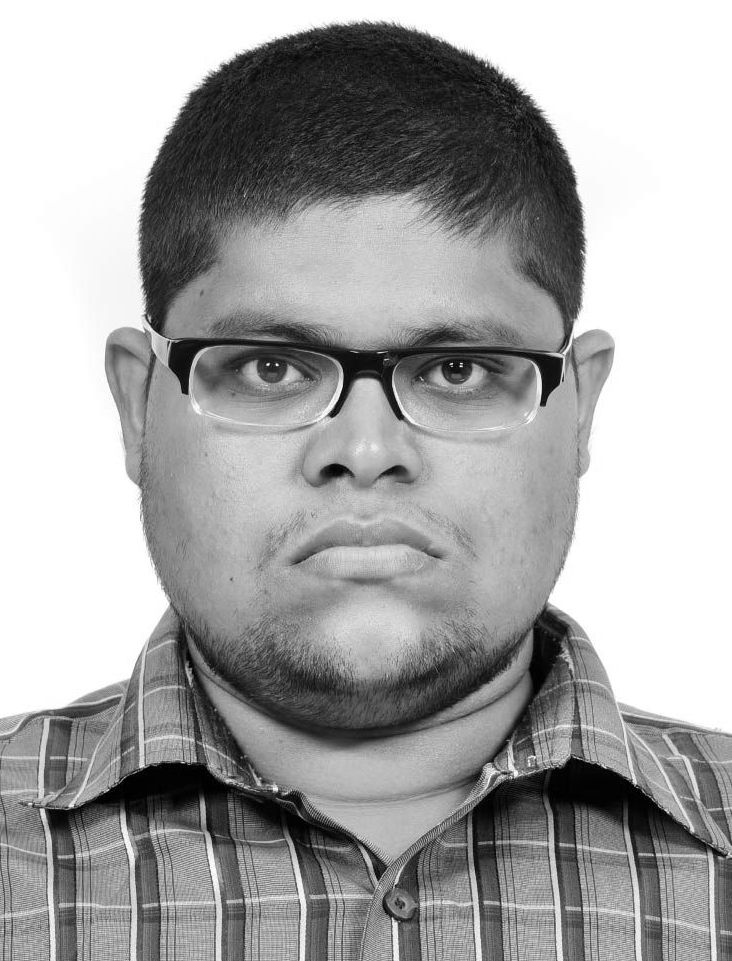}}]
{Chandranath Adak} (Senior Member, IEEE) received his PhD in Analytics from the University of Technology Sydney, Australia, in 2019. Currently, he is an assistant professor at the Dept. of CSE, IIT Patna, India. His research interests include Deep Learning, Data Analytics, and  Computer Vision.
\end{IEEEbiography}
\end{document}